\documentclass[eqsecnum,amsmath,preprintnumbers,superscriptaddress,nofootinbib,aps,11pt]
{revtex4} 
\pdfoutput=1
\usepackage{tikz}
\usetikzlibrary{calc,trees,positioning,arrows,chains,shapes.geometric,decorations.pathreplacing,decorations.pathmorphing,shapes,matrix,shapes.symbols}

\usepackage{bm}
\usepackage{MnSymbol}
\usepackage{amsmath}
\usepackage{graphicx}
\usepackage{feynmp}
\usepackage{hyperref}
\usepackage{bbold}
\usepackage{eufrak}
\usepackage{upgreek}
\usepackage{stmaryrd,scalerel}
\usepackage{enumitem}
\usepackage{color} 
\usepackage{etex}
\reserveinserts{18}
\usepackage{morefloats}

 \usepackage{longtable}

\newcommand{\cc}{\mathrm{g}}

\def\beq{\begin{equation}}
\def\eeq{\end{equation}}

\def\bea{\arraycolsep .1em \begin{eqnarray}}
\def\eea{\end{eqnarray}}
\def\Tr{{\rm Tr}}
\def\tr{{\rm tr}}

\def\!!!{\stackrel{!}{=}}

\def\a{\alpha}
\def\b{\beta}
\def\c{\gamma}

\newcommand{\g}{g}
\def\h{\eta}
\def\L{\Lambda}

\def\de{\delta}
\def\ge{\gamma}

\def\eps{\epsilon}

\def\Op{ \mathcal{F}}
\def\S{ \mathcal{S}}

\def\G{ \mathcal{G}}

\def\E{ \mathcal{E}}
\def\K{ \mathcal{K}}

\def\Det{{ \rm Det}}

\def\nn{ \nonumber \\}
\def\dL{\Lambda \partial_{\Lambda}}

\def\eq#1{(\ref{#1})}

\def\s0#1#2{\mbox{\small{$ \frac{#1}{#2} $}}}
\def\0#1#2{\frac{#1}{#2}}

\def\grgl{\:\hbox to -0.2pt{\lower2.5pt\hbox{$\sim$}\hss}{\raise3pt\hbox{$>$}}\:}
\def\klgl{\:\hbox to -0.2pt{\lower2.5pt\hbox{$\sim$}\hss}{\raise3pt\hbox{$<$}}\:}

\begin{document}
\title{ Background independent exact renormalisation}
\author{Kevin Falls\footnote{e-mail address: kfalls@sissa.it}}
\address{\mbox{Scuola Internazionale di Studi Superiori Avanzati (SISSA), via Bonomea 265, 34136 Trieste, Italy}}
\address{INFN, Sezione di Trieste, Italy}

\date{\today}

\begin{abstract}
A geometric formulation of Wilson's exact renormalisation group is presented based on a gauge invariant ultraviolet regularisation scheme without the introduction of a background field.
This allows for a manifestly background independent approach to quantum gravity and gauge theories in the continuum.
The regularisation is a geometric variant of Slavnov's scheme consisting of a modified action, which suppresses high momentum modes, supplemented by Pauli-Villars determinants in the path integral measure. An exact renormalisation group flow equation for the Wilsonian effective action is derived by requiring that the path integral is invariant under a change in the cutoff scale while preserving quasi-locality. 
The renormalisation group flow  is defined directly on the space of gauge invariant actions without the need to fix the gauge. We show that the one-loop beta function in Yang-Mills and the one-loop divergencies of General Relativity can be calculated without fixing the gauge. As a first non-perturbative application we find the form of the Yang-Mills beta function within a simple truncation of the 
Wilsonian effective action.

\end{abstract}

\maketitle

\newpage
\tableofcontents

\newpage

\section{Introduction}
Wilson's exact renormalisation group (ERG) \cite{Wilson:1973jj,Wilson:1975ly} can be used to investigate the non-perturbative structure quantum field theories.
The main idea is to solve an ERG flow equation for a scale dependent action where the microscopic action provides the initial condition.  As the effective scale is decreased the effects of integrating out high momentum fluctuations are encoded into the renormalised couplings. At the end point of the flow all fluctuations are integrated into the effective action allowing one to extract observables. 

Among the many applications of the ERG (see \cite{Berges:2000ew,Pawlowski:2005xe,Gies:2006wv,Rosten:2010vm} for reviews) are investigations into the phase structure of QCD \cite{Braun:2008pi,Braun:2009gm,Mitter:2014wpa,Fu:2016tey,Cyrol:2016tym,Cyrol:2017ewj,Fu:2019hdw} and studies of asymptotic safety in quantum gravity \cite{Codello:2007bd,Machado:2007ea,Benedetti:2009gn,Christiansen:2012rx,Falls:2013bv,Dona:2013qba,Codello:2013fpa,Falls:2014zba,Christiansen:2015rva,Meibohm:2015twa,Oda:2015sma,Ohta:2015efa,Demmel:2015oqa,Ohta:2015fcu,Dona:2015tnf,Denz:2016qks,Falls:2016msz,Gies:2016con,Knorr:2017fus,Knorr:2017mhu,Christiansen:2017bsy,Falls:2018ylp,Bosma:2019aiu,Knorr:2019atm,Burger:2019upn,Houthoff:2020zqy,Kurov:2020csd,Falls:2020qhj}. For these applications the gauge invariance of the theories 
 must be incorporated into the ERG formalism. Typically ERG studies are based on the effective average action approach \cite{Wetterich:1992yh,Morris:1993qb,Ellwanger:1993mw}.  For gauge theories a non-dynamical background field can be introduced \cite{Reuter1994181,Reuter:1996cp} to help control gauge invariance via the background field method.\footnote{ See \cite{Wetterich:2016ewc} for an alternative proposal  to the usual background field approach.}  
 The effective average action is then a gauge invariant functional of the background and dynamical field where gauge transformations act on both fields. Thus gauge invariance is gained at the price of an explicit background dependence.
 
 If we take gravity for example, the effective average action depends on a background metric $\bar{g}_{\mu\nu}$ and the dynamical metric $g_{\mu\nu}$.
As a consequence one has to keep track of all invariants which can be formed from the two metrics when solving the ERG equation for the effective action.  The dependence on the background field is governed by a split Ward identity  \cite{Pawlowski:2003sk,Pawlowski:2005xe,Safari:2015dva} which provides an extra constraint on solutions to the ERG flow equation. It is the Ward identity which ensures that observables are independent of the background field and invariant under physical gauge transformations which act on the dynamical fields alone. Thus, although seemingly paradoxical, background independence of observables is ensured by a non-trivial background dependence of the effective average action. Various works have  made use of the split Ward identities with applications in quantum gravity  \cite{Donkin:2012ud,Becker:2014qya,Dietz:2015owa,Morris:2016spn,Labus:2016lkh,Percacci:2016arh,Ohta:2017dsq,Nieto:2017ddk}, gauge theories \cite{Litim:2002ce} and scalar theories \cite{Bridle:2013sra,Safari:2016gtj}. However, within the background dependent approach to gauge theories and gravity, it remains a challenge to remove the dependence on the background field completely.

Clearly it would be advantageous if we could formulate the ERG in a manner which does not introduce a separate background field and preserves gauge invariance.
On the practical side removing the dependence on the background field reduces the number of independent couplings.
For example, if we consider terms which can be formed of two metrics without any spacetime derivatives there are already an infinite number of coupling constants \cite{Knorr:2017mhu}.  On the other hand if we have an action which is only a functional of the dynamical metric the only non-derivative coupling is the cosmological constant. Therefore there is a massive reduction in the number of couplings with an entire function $V(g,\bar{g})$ collapsing to a single term
\beq
\sqrt{\det \bar{g}} \, V(g_{\mu\nu},\bar{g}_{\mu\nu})  \to \lambda \, \sqrt{\det g} \,,
\eeq
with just a single free parameter $\lambda$, the cosmological constant.
Conceptually there is also the advantage that manifest background independence would bring the continuum approach closer to discrete background independent approaches to quantum gravity such as causal dynamical triangulations \cite{Loll:2019rdj}, Euclidean lattice theories \cite{Hamber:2015jja,Laiho:2016nlp} and tensor models \cite{Gurau:2016cjo}.

It therefore seems reasonable to ask whether one can avoid the dependence on the background field entirely by carrying out the renormalisation process in a manner which manifestly preserves gauge invariance. Indeed, it is perfectly possible to  regularise theories in a gauge invariant manner \cite{Slavnov:1977zf,Faddeev:1980be}. 
Manifestly gauge invariant formulations of the ERG which exploit this fact were first introduced in \cite{Morris:1995he} for pure $U(1)$ gauge theories.
This was generalised to $SU(N)$ Yang-Mills theories \cite{Morris:1998kz,Morris:1999px}, QCD \cite{Morris:2006in} and QED \cite{Arnone:2005vd}, while more recently these methods have been applied
classical gravity \cite{Morris:2016nda} and to gauge theories in curved spacetime \cite{Falls:2017nnu}. These formulation give the flow of the Wilsonian effective action as the ultraviolet (UV) cutoff scale is decreased and 
differ substantially for the flow of the effective average action which exploits an infrared (IR) regulator. Their advantage is that they are defined without gauge fixing and background fields  are never
introduced. However the form of the gauge invariant regularisation scheme \cite{Morris:2000fs,Morris:2000jj} which has been used so far utilises a spontaneously broken supersymmetric theory. This
introduces a set of regularising fields with interactions dictated by the supersymmetry in order to provide a necessary Pauli-Villars regulator.  As a  result the Wilsonian effective action has to depend on auxiliary dynamical
fields in addition to the dynamical fields of the original theory.

One can then ask whether it is possible to have a gauge invariant
flow equation for an effective action which depends only on the original set dynamical fields without introducing any additional fields, be they dynamical or non-dynamical. Again, this would seem to be a possible since the gauge invariant Pauli-Villars regularisation \cite{Slavnov:1977zf,Faddeev:1980be,Bakeyev:1996is} that is needed to regularise the path integral can be expressed simply as determinants of certain operators  without ever introducing additional fields.
Such a regularisation can be formulated in a geometric fashion by understanding the determinants as defining a regularised volume element on the orbit space of a gauge theory \cite{Asorey:1989ha}.

In this work we will  combine a geometric regularisation scheme with a similarly geometric version of the ERG which incorporate a non-trivial measure. In this manner we never have to introduce a background field or any additional auxiliary fields.  This approach will allow us to derive a flow equation for a Wilsonian effective action  for gauge theories and for gravity which is manifestly background independent. Since we will work in a gauge invariant setting we will not need to fix the gauge and therefore there will be no dependence on the gauge fixing parameters. Furthermore we will adopt a field covariant formalism that is reparameterisation invariant with the Wilsonian action transforming as a scalar.
The flow equation will therefore be of the same form whether we pick the fields to be metric $g_{\mu\nu}$, the inverse metric $g^{\mu\nu}$ or some rank two density such as $\sqrt{g} g_{\mu\nu}$.
 As a result we will not be afflicted with the gauge or parameterisation dependencies which are encountered in the effective average action approach \cite{Nink:2014yya,Falls:2015qga,Gies:2015tca,Percacci:2015wwa,Ohta:2016npm,Ohta:2016jvw}.

The remainder of the paper is organised as follows.
In section~\ref{secII} we construct the UV regularised path integral which will underlie the flow equation. In section~\ref{Ga_In} we  give a geometric discussion of gauge invariance and discuss various notions of locality for $n$-point functions. Then, in section~\ref{measure}, we derive the general form of the gauge invariant path integral. Along with the action $S[\phi]$ the path integral is parameterised by a choice of metrics $G$ and $H$ on the configuration space $\Phi$  and the gauge orbit $\G$ respectively. 
 In section~\ref{intgred} we introduce the main ingredients which we will assemble to construct the action and measure of the path integral. These include an action $I$, ultra-local metrics on $\Phi$ and $\G$ and differential operators which will appear in the a gauge invariant cutoffs. The explicit form of the regularisation is given in section~\ref{Reg_explicit} where we detail the form of the regularising terms in the action and the regularised path integral measure.    Section~\ref{secII} ends with a brief discussion and summary of the regularisation scheme. 

In section~\ref{secIII} we turn to the construction of the background independent flow equation beginning by stating the general criteria for such an equation. 
In section~\ref{Gen_Stru} we present the generalised structure of flow equations incorporating the measure. These flow equations are parameterised by the choice of two ERG kernels which specify the renormalisation scheme. A specific form of the ERG kernels which generates the gauge invariant regularisation is given in section~\ref{kernels}. We then demonstrate, in section~\ref{cutoff_sec}, that the functional trace appearing in the flow equation is cutoff in the UV. 

Section~\ref{secIV} demonstrates how standard one-loop results can be reproduced from the flow equation. In section~\ref{Scheme_Ind} we find the form of the on-shell one-loop effective action without specifying the scheme. The result is shown to agree with an expression  derived directly from the gauge fixed path integral.
 We then compute the one-loop renormalisation of the on-shell effective action in our scheme in section~\ref{one_loop_trace} showing that the logarithmic terms are universal. The one-loop beta function in Yang-Mills theory and the logarithmic divergencies of general relativity are reproduced. 
 
In section~\ref{Non-Pert_g} we compute the non-perturbative running of the Yang-Mills gauge coupling in a simple truncation of the effective action. 
 The form of the beta function is discussed and related to the breaking of scale invariance due to the measure in section~\ref{anom_dim_g}. We compare this calculation to the background field approximation of the effective average action in section~\ref{compare_to_BFA}.  
 
Our conclusions are presented in section~\ref{conc}.  

In appendix~\ref{OnShell_id} we collect some identities which hold when the equations of motion hold for a gauge theory. In appendix ~\ref{kappa_details}  we solve the classical flow up to terms which are quadratic in the equations of motion. Appendix~\ref{GFsec}  consider the gauge fixed path integral; in appendix~\ref{GF} we prove the gauge independence of the path integral while in appendix~\ref{GFflow} we show that our flow equation can be derived from the gauge fixed path integral.

\section{Geometric regularisation}
\label{secII}
 In this section we will construct a regularised path integral where we have in mind regularising a theory with a classical action $ I $ which is second order in spacetime derivatives. The action will then be supplemented by terms which regularise the theory at the momentum scale $\L$ and vanish for $\L \to \infty$ when the cutoff is removed. 
We will adopt the geometric approach to gauge theories \cite{Mottola:1995sj,Kunstatter:1991kw,dewitt2003global} by considering the space of all field configurations $\Phi$ as an infinite dimensional manifold.  The fields $\phi$ can then be thought of as coordinates on $\Phi$.  For gauge theories and gravity $\Phi$ has the structure of a fibre bundle where the fibres are the gauge orbits $\G$. Physics takes place on the quotient space $\Phi/\mathcal{G}$ where each point is associated to an equivalence class of field configurations under gauge transformations.  Explicitly we will be concerned with pure gravity and Yang-Mills theories. 
For gravity the fields are components of spacetime metric and the gauge transformations are diffeomorphisms whereas for Yang-Mills the fields are the gauge fields and the gauge transformations are local $SU(N)$ transformations. We will use the geometric approach to construct the path integral over $\Phi/\G$ in a manifestly gauge invariant manner without having to fix the gauge and thus we will never introduce any gauge breaking terms into the theory.
 Additionally using the geometric approach we can maintain field covariance such that we are free to  choose the field variables i.e. the coordinates on $\Phi$. 
  Here we will work exclusively with bosonic fields although a generalisation that includes fermionic fields is straightforward. We will also assume that spacetime has no boundaries such that we can freely integrate by parts and drop the boundary terms.

\subsection{Properties of $n$-point functions} 
\label{Ga_In}
Two important properties which we want to preserve are gauge invariance and quasi-locality. 
Here we will introduce our notation and explain how we can preserve gauge invariance by introducing gauge covariant tensors $T$ on $\Phi$ which correspond to gauge covariant $n$-point functions on spacetime. We will then explain the notations of locality, ultra-locality and quasi-locality.

\subsubsection{Gauge invariance}
In this paper we denote the fields by $\phi^a$ where latin indices from the start of the alphabet are DeWitt indices which include the spacetime coordinates $x$ as well as the discrete field indices. Thus for pure gravity when the field is taken to be the metric $\phi^a=g_{\mu\nu}(x)$ the DeWitt index $a$  includes the symmetric pair of lower Lorentz indices $\mu\nu$ (greek letters from the middle of the alphabet)  and the continuous spacetime coordinate $x$. However we are also free to use other coordinates on $\Phi$ transforming all $n$-point functions as tensors on $\Phi$. For example $\phi^a$ could also be chosen to a fluctuation $h_{\mu\nu}(x) = g_{\mu\nu}(x) - \bar{g}_{\mu\nu}(x)$ around some background or the its Fourier transform $\hat{h}_{\mu\nu}(p)$ in momentum space.  
We will denote an  infinitesimal gauge parameter by $\eps^\a$ with greek letters from the start of the alphabet understood as an additional set of DeWitt indices. In the case of gravity, where $\eps^\a= \eps^\mu(x)$ are vectors on spacetime, $\a$ includes an upper Lorentz index $\mu$ and the spacetime coordinate $x$.  The theory will be invariant under an infinitesimal  gauge transformation
  \beq \label{inf_gauge_transform}
\phi^a  \to \phi^a + K^a_\a[\phi] \eps^\a
\eeq
where $K^a_\a[\phi]$ are the generators of gauge transformation which are a set of vectors on $\Phi$.  As usual a repeated DeWitt index implies a sum over the discrete indices and an integral over spacetime.
The invariance of an action $S[\phi]$ is expressed by
\beq
S_{,a}[\phi] K^a_\a[\phi] = 0 
\eeq
where a comma followed by a latin DeWitt index denotes a functional derivative with respect to $\phi^a$. 
If we take the explicit example of gravity the transformation \eq{inf_gauge_transform} takes the form 
\beq
g_{\mu\nu}(x)  \to g_{\mu\nu}(x) +  \eps^{\rho}(x) \partial_{\rho} g_{\mu\nu}(x) +g_{\mu\rho} \partial_{\nu} \eps^{\rho}(x)   +  g_{\nu\rho}(x) \partial_{\mu} \eps^{\rho}(x) \,,
\eeq 
such that explicitly 
\beq
K^a_\a =  \delta(x,y)  \frac{\partial}{\partial x^{\rho}} g_{\mu\nu}(x)  +   g_{\mu\rho}   \frac{\partial}{\partial x^{\nu}}    \delta(x,y) +  g_{\rho\nu}  \frac{\partial}{\partial x^{\mu}}   \delta(x,y)\,.
\eeq
More generally we assume that the gauge algebra closes off-shell such that the Lie bracket of two generators is given by
\beq \label{Lie_Algebra}
[\vec{K}_{\alpha}, \vec{K}_{\beta}] = f^{\gamma}\,_{\alpha \beta} \vec{K}_{\gamma}\,,
\eeq
which can be expressed in component form as
$
\mathcal{L}_{K_\a} K^a_\b \equiv   K^b_{\alpha} K_{\beta,b}^a - K^b_{\b} K_{\a,b}^a= f^{\gamma}\,_{\alpha \beta} K^a_{\gamma}\,.
$
Here $f^{\alpha}\,_{\beta \gamma}= - f^{\alpha}\,_{\c \b} $ are the structure constants with $f^{\alpha}\,_{\beta \gamma,a}=0$. The property \eq{Lie_Algebra}  holds in particular for $SU(N)$ gauge theories and for gravity which are the cases of interest in this paper. For both gravity and $SU(N)$ theories $K^a_\a$ is a first order differential operator. 
Additionally for certain choices of the fields variables $\phi^a$ one has
\beq \label{linear}
K^a_{\a}[\phi] \!!! k^a_{\a b} \phi^b + k^a_{\alpha}\,,
\eeq
with $k^a_{\a b,c}= 0 = k^a_{\alpha,b}$. Although this property does not hold for all choices of parameterisations $\phi^a$.

Since physics takes place on $\Phi/\G$ we will
work with tensors  $T^{a_1 a_2 ...}\,_{b_1 b_2 ...}[\phi]$ on $\Phi$ which transform covariantly under gauge transformations. In particular by forming a scalar on $\Phi$ from such covariant tensors we can construct gauge invariants such as terms in the action $S[\phi]$. The {\it gauge covariance} of a tensor (density) on $\Phi$ can be expressed as
\beq \label{gauge_covariant_field-space_tensor_densities}
\mathcal{L}_{K_{\alpha} } T^{a_1... a_n }\,_{b_1... b_n}[\phi] = 0\,,
\eeq 
where $\mathcal{L}_{K_{\alpha}}$ denotes the Lie derivative with respect to the generators $K_{\a}$ which we view as vectors on $\Phi$.
In addition to covariant tensors, which carry only latin  $\Phi$-indices, we will work with tensors on $\Phi$ which carry greek $\G$-indices as well (e.g. $K^a_{\a}$).
The gauge covariance of these tensors implies the more general identity
\bea \label{Ward}
&&\mathcal{L}_{K_{\alpha} } T^{a_1... a_n \a_1...  \a_i }_{  \b_1...\b_j \, b_ 1... b_m }[\phi]  -  T^{a_1... a_n  \gamma...  \a_i }_{\b_1...\b_j\, b_ 1...b_m  }[\phi]   f^{\a_1}\,_{\gamma \a}    
 -  ...
 +  T^{a_1... a_n  \a_1...  \a_i }_{\c...\b_j \,\, b_ 1...b_m  }[\phi]f^{\c}\,_{\b_1 \a} 
 + ...  = 0 \,,
\eea  
which can be understood as the most general form of a Ward identity for a given covariant tensor $T[\phi]$ on $\Phi$ which also carries gauge orbit indices.
It is easy to see that $K^a_\a$ is covariant as a consequence of \eq{Lie_Algebra} and that the covariance of $f^\a\,_{\b\c}$ implies the Jacobi identity. One can show that the vertices of a gauge invariant action $S_{,a_1..a_n}$ are covariant provided that \eq{linear} holds for the fields.

To avoid expressions involving delta functions it is useful to express $n$-point functions, such as $K^a_\a$, by introducing test functions $\delta \phi^a$, $ \delta J_a$, to contract indices free latin indices, and test functions $\delta \xi ^{\a}$ and $\delta j_{\a}$, to contract free greek indices. When the $n$-point function has multiple indices of the same type one can label distinct test fields e.g. $T_{ab} \delta \phi_1^a \delta \phi_2^b$ (although if the tensor is symmetric in two or more indices this is unnecessary for those indices). The tensor is then expressed as a functional $T[\phi,\delta \phi_i, \delta J_j, \delta \xi_k , \delta j_l] $ of the fields $\phi$ and the test fields. e.g. 
\beq
K[\phi, \delta J_a , \delta \xi^a ] = \delta J_a K^a_{\a}[\phi] \delta \xi^\a \,.
\eeq
Then the functional is invariant under a transformation \eq{inf_gauge_transform} provided the test fields transform as
\bea
\delta \phi^a \to \delta \phi^a +  K^a_{\a ,b} \delta \phi^b \eps^\a  \,,\\
\delta J_a \to  \delta J_a - K^b_{\a ,a} \delta J_b \eps^\a \,,\\
\label{deltaxi_transform}
\delta \xi^\a  \to  \delta  \xi^\a + f^\a\,_{\b \c} \delta \xi^\b \eps^\c \,, \\
\delta j_\a  \to  \delta  j_\a - f^\b\,_{\a \c} \delta j_\b \eps^\c \,.
\eea
To understand the origin of the transformation of $\delta \xi^\a$ given by \eq{deltaxi_transform} consider a perturbation of the field which is purely  gauge
\beq
\delta \phi^a = K^a_\a  \delta \xi^\a
\eeq
 then on one hand the transformation of $\delta \phi$ implies
\beq
\delta \phi^a \to \delta \phi^a  +    K^a_{\a ,b}    K^b_\b  \delta \xi^\b \eps^\a
\eeq
while on the hand  the transform of $K^a_\a  \delta \xi^\a$ gives
\beq
\delta \phi^a \to \delta \phi^a  +    K^a_{\a,b}  \delta \xi^\a K^b_\b  \eps^\b + K^a_\a  f^\a_{\b\c} \delta \xi^\b \eps^\c
\eeq
but these to expressions agree by the definition of the the structure constants \eq{Lie_Algebra}.
A dictionary between DeWitt notation and explicit notation for Yang-Mills and gravity is given in table~\ref{tab_DeWitt}.

\begin{center}
\begin{table}
\begin{tabular}{ | c | c | c | }
    \cline{1-3}\noalign{\smallskip}
 DeWitt&Yang-Mills & Gravity\\
  \noalign{\smallskip}\hline\noalign{\smallskip}
    $\phi^a$ & $A_\mu^i(x)$& $g_{\mu\nu}(x)   $ \\
      \noalign{\smallskip}\hline\noalign{\smallskip}
    $\delta\xi^\a$ & $\delta\omega^i(x)$& $\delta v^\mu(x)   $ \\
      \noalign{\smallskip}\hline\noalign{\smallskip}
    $S_{,a}$ & $  \frac{\delta S[A]}{\delta A_\mu^i(x)  }$& $  \frac{\delta S[g]}{\delta g_{\mu \nu }(x)  }   $ \\
      \noalign{\smallskip}\hline\noalign{\smallskip}
$K^a_{\a} \delta \xi^\a$ & $\nabla_{\mu} \delta \omega^j(x) $& $ \nabla_{\mu} \delta v_{\nu}(x) +  \nabla_{\nu} \delta v_{\mu}(x)    $ \\
      \noalign{\smallskip}\hline\noalign{\smallskip}
       $K^a_{\a} \delta J_{a} $ & $\nabla_{\mu}  \delta J^{\mu}_i(x)$      &  $ 2 \nabla_{\mu}  \delta \tilde{J}^{\mu}\,_\nu(x) $ \\
      \noalign{\smallskip}\hline\noalign{\smallskip}
$f^{\a}_{\b\c} \delta\xi_1^{\b}  \delta\xi_2^{\c} $ &   $ f_{ijk} \delta \omega_{1}^j(x) \delta\omega_{2}^k(x) $  & $\mathcal{L}_{\delta v_2}\delta v^{\mu}_1(x) $\\
  \noalign{\smallskip} \hline
\end{tabular}
\caption{DeWitt notation is summarised with several examples. In the first column we give the DeWitt notation with the corresponding object expressed in standard notation for Yang-Mills and Gravity appearing in the same row.  Here $\nabla$ denotes the covariant derivative with the $A_{\mu}$ being the connection for Yang-Mills and the Levi-Cevita connection for gravity. We note that for gravity the components of the dual basis to $\delta g_{\mu\nu}(x)$ are tensor densities $\delta \tilde{J}^{\mu\nu}(x)$ of weight one as indicated by the tilde.  }
\label{tab_DeWitt}
\end{table}
\end{center}

\subsubsection{Locality}
Viewing $n$-point functions as functionals of the fields $\phi$ and a set of $n$ test fields  $T[\phi,\delta \phi_i, \delta J_j, \delta \xi_k ,\delta j_l] $ one can generalise notions of locality for local action functionals (i.e. $0$-point functions) to arbitrary $n$-point functions.
 Locality is the property that we can express the functional  $T[\phi,\delta \phi_i, \delta J_j, \delta \xi_k ,\delta j_l] $ as a spacetime integral over a function of the fields and their derivatives  such that
\beq
T = \int d^dx \, \mathcal{T}(\phi(x), \partial \phi(x), ...,  \delta \phi_i,    \partial \delta \phi_i,..., \delta J_j  , \partial   \delta J_j,..., \delta \xi_k, \partial \delta \xi_k ...,\delta j_l, \partial \delta j_l,...  )  \,.
\eeq
Note that it follows that by taking functional derivatives of a local functionals we obtain local functionals.
Ultra-locality is the special case where  $\mathcal{T}$ is only a function of the fields with no derivatives acting on the any of the fields. 
Quasi-locality corresponds to a case where $T$ is non-local but where we can expand $T$ in a derivative expansion 
\beq
T \simeq \int d^dx \,\sum_{m=0}^{\infty} \mathcal{T}_m(\phi(x), \partial \phi(x), ...,  \delta \phi_i,    \partial \delta \phi_i,..., \delta J_j  , \partial  \delta J_j,..., \delta \xi_k, \partial \delta \xi _k ...,\delta j_l, \partial \delta j_l,...  )
\eeq
where each term $ \mathcal{T}_m$ in the expansion involves $m$ derivatives. 

The importance of quasi-locality is that it ensures that there are no IR divergencies since in momentum space quasi-locality states that $n$-point functions have an analytic expansion around vanishing momentum.
In this work we will insist that all $n$-point functions which enter the regularisation and the flow equation are quasi-local in order to be sure not to encounter any IR singularities.

\subsection{Gauge invariant measure}
\label{measure}

The regularised functional integral will take the form
\beq \label{ZGaugeInv}
\mathcal{Z} =     \int_{\Phi/\G} d \{\phi\} \, \tilde{M}[\{\phi\}] \, e^{-S[ \{\phi \} ]}
\eeq
where the integral is over the space of orbits $\Phi/\G$  with $\{\phi\}$ denoting equivalence classes of field configurations under gauge transformations  and  $\tilde{M}[\{\phi\}]$ denotes a volume element on 
$\Phi/\G$.
To give a definite meaning to the measure \cite{Mottola:1995sj,Falls:2017cze} we introduce a metric $G_{ab}[\phi]$ on $\Phi$ as well as a metric $H_{\a\b}[\phi]$ which we use to define a Haar measure on $\G$.
The measure   $d \{\phi\} \, \tilde{M}[\{\phi\}]$  on $\Phi/\G$  is then defined by integrating over $\Phi$ with the measure $d \phi  \sqrt{\det G[\phi]} $ while extracting an integral over the gauge group such that 
\beq \label{Measures}
 d \phi  \sqrt{\det G[\phi]}    =   d \{\phi\}   \tilde{M}[\{\phi\}]   \, d \xi    \sqrt{\det H[\phi]}   \, .
\eeq  
With this definition of $\tilde{M}[\{\phi\}]$ the path integral can be written as
\beq \label{pathintegral}
\mathcal{Z} = \mathcal{N} \int_\Phi d\phi  \frac{\sqrt{\det G[\phi] }}{\sqrt{\det H[\phi] }}   e^{-S[\phi ]  } \,.
\eeq
where $\mathcal{N}= 1/\int_\G d\xi$.


Although we will not need to fix the gauge in this work let us demonstrate that the path integral \eq{pathintegral} is equal to the BRST invariant path integral obtained when we perform the Fadeev-Popov trick. In particular let us check that  the definition of the measure \eq{Measures} leads to BRST invariant measure after gauge fixing. To do so we consider a surface in $\Phi$ defined by
\beq \label{gauge_condition}
\chi^\a[\phi] = \omega^\a\,.
\eeq
Then we write a one in the form
\beq
1 =   \int_\G d \xi   \delta(\chi[\phi_\xi ] - \omega)      \det Q[\phi_\xi]  
\eeq
where $Q^\a\,_\b[\phi] =  \det \chi^\a_{,a}[\phi] K^a_\b[\phi] $
and $\phi_\xi$ denotes the transformed field.
Inserting the $1$ into \eq{ZGaugeInv} and using \eq{Measures} yields
\bea
\mathcal{Z} 
&=&      \int_{\Phi} d\phi  \frac{\sqrt{\det G[\phi] }}{\sqrt{\det H[\phi] }} \, e^{-S[\phi ]}      \delta(\chi[\phi] - \omega)       \det Q[\phi]
\eea
Integrating over $\omega$ with a Gaussian weight $e^{-\frac{1}{2} \omega \bar{Y} \omega }$ and introducing the usual ghost fields we can then write
\beq
\mathcal{Z} =     \int_\Phi d\phi  \sqrt{\det G[\phi] }  \int d C_{\rm gh} \frac{1}{\sqrt{\det H[\phi]} }  \int d\bar{C}_{\rm gh}  \int d B_{\rm gh}  e^{-S[\phi] - S_{\rm BRST}[\phi,C_{\rm gh}, \bar{C}_{\rm gh},  B_{\rm gh}]  }
\eeq
where we introduce the BRST invariant action
 \beq
 S_{\rm BRST}[\phi,C_{\rm gh}, \bar{C}_{\rm gh},  B_{\rm gh}] =  \frac{1}{2} B_{\rm gh}^{\a} \bar{Y}_{\a\b} B_{\rm gh}^\b + i B_{\rm gh}^\a \bar{Y}_{\a\b} \chi^\b + \bar{C}_{\rm gh}^\a \bar{Y}_{\a\b} Q^{\b}\,_{\c} C^{\c}_{\rm gh}
 \eeq
 where BRST transformation is given by
 \bea
\delta_\theta \phi^a& =& K^a_{\alpha} C^{\alpha}_{\rm gh} \theta \,, \\
\delta_\theta C^{\alpha}_{\rm gh} &=& \frac{1}{2}C^{\beta}_{\rm gh} f^{\alpha}_{\beta \gamma}  C^{\gamma}_{\rm gh} \theta \,, \\
\delta_\theta \bar{C}^{\alpha}_{\rm gh} &=&  i B^{\a}_{\rm gh} \theta \,,
\\
\delta_\theta B^{\bar{\alpha}} &=& 0 \,,
\eea
with $\theta$ an anti-commuting parameter. Then we observe that the measure is also BRST invariant with the factor  $(\det H[\phi])^{-1/2}$ providing the right volume element to keep the $C_{\rm gh}$-integral BRST invariant. A proof of gauge independence is given in appendix~\ref{GF}.

\subsection{Basic ingredients}
\label{intgred}

Now we will introduce `basic ingredients' which will be assembled to construct the regularisation of the theory and the corresponding flow equation.
These will consist of an action $I$, an ultra-local metric on $\Phi$, denoted $\gamma_{ab}$, an ultra-local metric on $\G$, denoted $\eta_{\a \b}$, and the generators $K^{a}_{\a}$ which we have introduced previously. All other $n$-point functions which will be introduced will be `decedents' of these basic ingredients obtained by taking functional derivatives and contracting DeWitt indices. 
 We will assume that the action functional $I$  is second order in derivatives such that it includes a canonical kinetic term for the theory. For gravity the action is given by the Einstein-Hilbert action
\beq \label{EH}
 I[\phi] = - \frac{1}{16 \pi G_N} \int d^Dx \sqrt{g} (R - 2 \bar{\lambda}) \,,
 \eeq
 where $\sqrt{g} \equiv \sqrt{\det g} $,  $G_N$ is Newton's coupling, which appears in the coefficient of the Einstein-Hilbert term in $S$, and we have also allowed for a non-zero cosmological constant $\bar{\lambda}$, which is optional. 
 For $SU(N)$ Yang-Mills the action is given by
 \beq
I[\phi] = \frac{1}{4\cc^2} \int d^Dx F_{i \mu\nu} F^{i \mu\nu}\,,
\eeq
 where  $\cc$ is the gauge coupling.
Generally $I$ will be proportional to a coupling $Z(\L)$ which plays the role of the wave-function renormalisation. For the Einstein-Hilbert action $Z(\L) = 1/(32 \pi G_N)$ while for a gauge theory $Z(\L) = 1/ {\rm g}^2$. We define the anomalous dimension as 
 \beq
 \upeta(\L) \equiv   - Z(\L)^{-1} \dL Z(\L)\,.
 \eeq
  We also introduce an ultra-local metric $\gamma_{ab}$ which should have dimensionless line element $\delta \phi^a \gamma_{ab} \delta \phi^b$. For Einstein gravity we take the ultra-local metric to be given by the DeWitt metric
\beq \label{gamma}
\ge_{ab} \delta \phi^a \delta \phi^b  =    \frac{\Lambda^{2}}{32 \pi G_N} \frac{1}{2}  \int d^Dx \sqrt{g} \left(\g^{\mu\rho}\g^{\nu \sigma} + \g^{\mu\sigma}\g^{\nu\rho} -  a \g^{\mu\nu} \g^{\rho \sigma} \right) \delta \g_{\mu\nu} \delta \g_{\rho\sigma}\,
\eeq 
where $a$ is the DeWitt parameter  which we shall fix here to be
\beq \label{DeWittParameter}
a=1
\eeq
although other choices may be possible.
Similarly we introduce an ultra-local metric $\eta_{\a\b}$ on $\G$ which should have a dimensionless line element $\delta \xi^{\a} \eta_{\a\b} \delta \xi^\b$. For Einstein gravity we take this metric to be given by  the line element
\beq
\eta_{\alpha \beta}  \delta\xi^{\alpha} \delta\xi^{\beta} =   \frac{ \Lambda^{4}}{16 \pi G_N} \int d^Dx \sqrt{\g} \g_{\mu\nu}  \delta v^{\mu} \delta v^{\nu} \,. 
\eeq
For the case of $SU(N)$ theory the metrics are given by
 \beq
\ge_{ab} \delta \phi^a \delta \phi^b  =\frac{\Lambda^2}{\cc^2}  \int d^Dx \sqrt{g} g^{\mu\nu} \delta_{ij} \delta A^{i}_{\mu} \delta A^{j}_{\nu} \,,
 \eeq 
 and
 \beq
\h_{\a\b} \delta \xi^\a  \delta \xi^\b  =  \frac{\Lambda^4}{\cc^2}  \int d^Dx \sqrt{g}  \delta_{ij}  \delta \omega^{i} \delta \omega^{j}\,.
 \eeq  
where $\delta_{ij}$ is the Cartan-Killing metric which is simply a Kronecker-delta in the case of $SU(N)$ Yang-Mills. 

Note that the powers of $\L$ appearing in $\gamma_{ab}$ and $\eta_{\a\b}$ ensure that the line-elements are dimensionless. 
Generally for two derivative actions such as Yang-Mills and Einstein-Gravity we have that $\gamma_{ab} \propto \L^2 Z(\L)$ and $\eta \propto \L^4 Z(\L)$ which implies that
\beq
\dL \gamma_{ab} = (2 - \upeta) \gamma_{ab}\,,      \,\,\,\,\,\,\,\,\,\,    \dL \eta_{\a\b} = (4 - \upeta) \eta_{\a\b} 
\eeq 
The inverse metrics of $\gamma_{ab}$ and $\eta_{\a\b}$ will be denote by $\gamma^{ab}$ and $\eta^{\a \b}$ respectively where we will use $\gamma$ and $\eta$ to raise and lower indices  e.g.  
 \beq
 X^a\,_b \equiv X^{ac} \gamma_{cb},   \,\,\,\,\,\, Y^\a\,_\b = Y^{\a\c} \eta_{\c \b}\,.
 \eeq
 
 The metric $\gamma_{ab}$ allows us to define a covariant derivative $\nabla_a$ with  the Levi-Civita  connection constructed from $\gamma_{ab}$ such that $\nabla_c \gamma_{ab} = 0$.
 We will also use $;$ in the subscript (in place of $,$) to denote a covariant functional derivative   and the notation $S^{(n)}$ to denote the $n$th covariant functional derivative of a functional.   For the DeWitt metric \eq{gamma} the corresponding Levi-Civita connection is given by
\beq
\Gamma^c_{ab}  \delta J_c  \delta\phi^a \delta \phi^b = \int d^Dx \left( -\delta^{\mu}_\psi  \delta^{\sigma}_{\omega} g^{\nu \rho}   + \frac{1}{2} g^{\mu\nu} \delta^{\rho}_{\psi} \delta^{\sigma}_{\omega}  + \frac{g_{\psi \omega}}{4 (D-2)} (2 g^{\mu \rho} g^{\nu \sigma}   -  g^{\mu\nu} g^{\rho \sigma} ) \right)  \delta g_{\mu \nu} \delta g_{\rho \sigma} \delta \tilde{J}^{\psi \omega}  \nonumber \,,
\eeq 
while for Yang-Mills the connection is flat.

Let's note that the mass dimension
 of each of the basic ingredients and there derivatives follows from their tensor structure: the dimension $[T]$ of a tensor 
$T$ is given by  $[T] = (\#{\rm \,of\,upper\,latin\,indices} - \#{\rm \,of\,lower\,latin\,indices})[\phi]  + (\#{\rm \,of\,upper\,greek\,indices}-\#{\rm \,of\,lower\,greek\,indices}) [\xi]$.
It follows that two point functions obtained from the basic objects   $I$, $\gamma$, $\eta$ and the gauge generators $K^a_{\a}$ with one raised index and one lower index of the same type, e.g. $X^a\,_b$ and $Y^{\a}\,_\b$, will be dimensionless.

A two point function of this type is the dimensionless second order differential operator given by 
\beq \label{Delta}
\Delta^{a}\,_b \equiv \gamma^{ac} I_{;cb} + K^a_{\a}\eta^{\a\b} K_{\b}^c\gamma_{cb}\,,
\eeq
where due to the second term $\Delta^{a}\,_b$ is invertible.
For gravity the operator $\Delta$ is given by
\bea \label{DeltaGrav}
\delta J_a \Delta^a\,_b \delta \phi^b &=& \int d^Dx \frac{\delta \tilde{J}^{\mu\nu}}{\L^2}   \Bigg(  - \nabla^2 \delta_\mu^\rho  \delta_\nu^\lambda    - R \frac{1}{2 (D-2) } g_{\mu\nu} g^{\rho \lambda}   + \frac{1}{2} R_{\mu\nu} g^{\rho \lambda}    + \frac{1}{D-2} R^{\rho \lambda} g_{\mu\nu} \nn
&&+ \frac{(D-2) R-2 D \bar{\lambda} }{2 (D-2)} \delta_\mu^\rho  \delta_\nu^\lambda  - 2 R^{\rho}\,_\mu\,^\lambda\,_{\nu}  \Bigg) \delta g_{\rho\lambda} \,,
\eea
where as for Yang-Mills
\beq \label{DeltaYM}
 \delta J_a \Delta^a\,_b \delta \phi^b     =  \int d^Dx \delta J_{i}^{\mu} \Lambda^{-2} (-\nabla^2\delta A_{\mu}^i  + 2F_{j\mu}\,^{\nu} f^{ji}\,_{k} \delta A_{\nu}^k  ) 
\eeq
For gravity $\nabla= \nabla_g$ is the covariant derivative with the  Levi-Civita connection for $g$ while for Yang-Mills $\nabla=\nabla_A$ denotes the gauge covariant derivative where $A_\mu^i$ is the connection. The eigenvalues $p^2/\Lambda^2$ of $\Delta^{a}\,_b$ can then be thought of as momenta in units of $\L^2$.

Given $\gamma_{ab}$ there is also a natural  operator  $\Delta^\parallel_{\a\b}$  formed by the projection of $\gamma_{ab}$ onto $\G$ 
\beq \label{Delta_K}
\Delta^\parallel_{\a\b}= K^{a}_{\a} \gamma_{ab} K^{b}_{\b}
\eeq
 which for gravity is given by
\beq
\Delta^\parallel_{\a\b} \delta \xi^\a \delta \xi^\b  = \frac{  \Lambda^2}{16 \pi G_N}  \int d^Dx \sqrt{\det g} \,  \delta v^\mu (- g_{\mu\nu}  \nabla^2   - R_{\mu\nu})  \delta v^\nu\,.
\eeq
By raising one index with $\eta^{\a\c}$  we can then form a dimensionless operator
$
(\Delta^\parallel)^\a\,_\b \equiv    \eta^{\a\c} \Delta^\parallel_{\c\b} 
$
 whose eigenvalues $p^2/\L^2$ we can think of as the defining the ghost momenta in units of $\L^2$.
 For gravity we have
 \beq
 (\Delta^\parallel)^{\mu}\,_\nu = - \Lambda^{-2} (\nabla^2 \delta^\mu\,_\nu + R^\mu\,_\nu) \,,
 \eeq
 and for Yang-Mills we have
 \beq \label{DeltaghYM}
  (\Delta^\parallel)^{i}\,_j = - \Lambda^{-2} \nabla^2 \delta^i_j  \,.
 \eeq
 We can define the operator
\beq
(\Delta^\perp)^{a}\,_b \equiv \gamma^{ac} I_{;cb} \,,
\eeq
which is transverse when $I_{,a} = 0$ and hence cannot be inverted. 
It is also useful to define projection operators 
\beq
\Pi^a\,_b = \delta^a_b - \Xi^a\,_b\,,   \,\,\,\,\,\,\, \,\,\,\,\,\,\,\,\,\,\,  \Xi^a\,_b = K^a_{\alpha} \mathfrak{G}^{\a\b} K_{\beta}^c\gamma_{cb}\,,
\eeq
where $\mathfrak{G}^{\a\b}$ is the Greens function 
\beq
\Delta^\parallel_{\a\c} \mathfrak{G}^{\c\b} = \delta^\b_\a \,,
\eeq
and where $\delta^\a_\b$ and $\delta^a_b$ are the identities.  
Due to the presence of the non-local operator $\mathfrak{G}^{\a\b} $ one should generally avoid regularisation schemes which depend on the projection operators since this would be a source of potential IR singularities.
However inserting $\delta^a_b = \Pi^a\,_b  + \Xi^a\,_b$ into equations is useful calculation method which can be exploited.
In appendix~\ref{OnShell_id} we derive several useful identities which apply when the field equations 
\beq \label{EofM}
\E_{a} \equiv I_{,a} = 0\,,
\eeq
are satisfied.

 \subsection{Regularisation}
 \label{Reg_explicit}
 Let us now turn to the choice of regularisation using the basic ingredients which we specified in the previous subsection. 
 The idea is to use $I$, $\eta$, $\gamma$ and $K$ to construct the metrics $G$ and $H$ in addition to a UV regulator term $S_{\rm UV}$ in the action $S$. 

 First we recall that one can improve the UV behaviour of a gauge theory by adding terms, i.e. the UV regulator $S_{\rm UV}$, to the action which should vanish in the limit $\L \to \infty$ when the cutoff is removed. These terms will involve higher order covariant derivatives suppressed by $\L^{-1}$ ensuring that the propagator of the theory (in the transverse space)  falls to zero for high momentum \cite{Slavnov:1972sq,Lee:1972fj}. One can choose $S_{\rm UV}$ such that no extra poles are introduced into the propagators if one allows the terms to be quasi-local (i.e. involving an infinite number of derivatives) \cite{Tomboulis:1997gg, Modesto:2011kw}.
Provided the propagator decreases quickly enough the theory will only have a divergences at one-loop and in multi-loop diagrams that contain the divergent one-loop diagrams as sub-diagrams. Such a theory is super-renormalisable only requiring counter terms at one-loop. 
To make the theory finite the path integral measure must also be modified in such a way as to regularise also the one-loop diagrams. This can be achieved by including gauge invariant determinants of Pauli-Villars (PV) operators in the measure \cite{Slavnov:1977zf} which reduce to the identity for low momenta or when $\L \to \infty$. Here we shall include these operators in the measure by choosing $G$ and $H$ to be products of PV operators. When $\L \to \infty $ we will demand that $G \to \gamma$ and $H \to \eta$ and $S_{\rm UV} \to 0$.

Ignoring gauge invariance momentarily, a naive choice is to set 
\beq \label{G_not_good}
G_{ab} =\gamma_{ab} + S^{\rm UV}_{;ab}
\eeq
since $ S^{\rm UV}_{;ab}$ diverges strongly in the high momentum limit $G_{ab}\sim S_{,ab}$. The functional trace $\Tr \log \left( G^{-1}  S^{(2)} )  \right)$ is then finite provided the high momentum limit is reached fast enough. The term $\gamma_{ab} \propto \L^2 $ in \eq{G_not_good} is then seen to be a Pauli-Villars mass matrix.
However,  the vertices of the PV operators, obtained by taking derivatives of \eq{G_not_good} will diverge for high momentum at the same rate as the two point function $S_{,ab}$ and this introduces further unregulated `overlapping' divergences.  
 Thus for the choice \eq{G_not_good} we will generically have new divergent diagrams which come from expanding $\sqrt{\det G }$.
For example, consider a diagram with a singe physical loop and external PV lines, if the vertices diverge at the same rate as the inverse propagator these diagrams will be divergent.   

The introduction of overlapping divergences is avoided if the propagators for the physical fields decrease sufficiently faster than the vertices of Pauli-Villars fields diverge \cite{Faddeev:1980be}.
Thus instead of including only one Pauli-Villars operator in $G$ we should choose $G$ such that it factors into operators which are less divergent. Specifically, {\it instead} of \eq{G_not_good},  we will take $G$ to be the square of such Pauli-Villars operators 
 \beq \label{G}
G_{ab} = (C^{-1})^{c}\,_a \gamma_{cd} (C^{-1})^d\,_b 
\eeq
then to regularise the one-loop diagrams we must have that that  $C^{2} S^{(2)} \sim 1$ at high momentum. 
 In this way the rate at which the vertices of $C^{-1}$ diverge is halved  and no new divergent diagrams will be introduced by the measure. 
 Consistent regularisation schemes which avoid the overlapping divergences in this manner can then be constructed \cite{Asorey:1989ha,Bakeyev:1996is}.
 Below we will specify the form of our regularisation based on the choice \eq{G}. Note that since $S^{(2)}$ diverges in the high momentum limit $C$ will behaves as a UV cut-off vanishing at high momentum.

 For a gauge theory we actually require that $C$ must regulate the transverse fluctuations meaning that for high momentum
\beq  \label{G_condition}
\Pi^{c}\,_a G_{cd} \Pi^{d}\,_b   \sim   S_{;ab} \,.
\eeq
Although $S_{;bc}$ is not transverse (i.e. it does not commute with $\Pi$) it will be transverse for high momentum since the projection onto longitudinal modes $S_{;ab} K^b_{\a} =- S_{,b} K^b_{\a;a}$ is a first order differential operator and thus  only diverges with one power of momentum. 
While \eq{G_condition} is a restriction on the transverse projection of $G$ being the PV regulator for the physical fluctuations there remains longitudinal projection of $G$ which must also be regulated.
This is achieved by demanding that
 \beq \label{H_condition}
  K^a_{\a} G_{ab} K^b_\b  \sim H_{\a\b} \,,
 \eeq
which takes into account the Jacobian when encountered when factoring out the gauge volume from the functional measure. Thus we can think of $H$ as providing the PV regulator for the ghosts.
 To satisfy these requirements we choose $H$ to be given by the product
\beq \label{H}
H_{\a\b} = B^{\c}\,_{\a} b_{\c \eps} B^{\eps}\,_\b
\eeq
where 
\beq \label{B_condition}
( C^{-1})^a\,_b K^b_\a \sim K^{a}_\b B^{\b}\,_\a 
\eeq
ensures that the longitudinal  part of $\det C$ is regulated by $\det B$. Then to satisfy \eq{H_condition} the require that for high momentum
\beq \label{b_condition}
b_{\a \b} \sim \Delta^\parallel_{\a\b} = K^a_{\a} \gamma_{ab} K^b_\b\,.
\eeq
For low momentum we can require that $C$, $B$ and $b$ go to the identity such that the in the continuum limit $\L \to \infty$ we have $G = \gamma$ and $H = \eta$.

 Let us now introduce the action
 \beq
 A = I + \frac{1}{2} I_{,a} X^{ab} I_{,b}
 \eeq
 where $X^a\,_b = (X(\Delta))^a\,_b$ is a function of $\Delta$. It is convenient to parameterise the choice of $X(z)$ in terms of a cutoff function $c(z)$ where
 \beq	 \label{c_0}
 X(z) = \frac{1- c(z)}{z c(z)}
 \eeq 
and the cutoff has the limits $c(0) = 1$ and $c(\infty)=0$ and assume $c(z)$ to be monotonic and have an analytic expansion around $z=0$. 
 It follows that $c(z)$ has no zeros for $z>0$ along the real axis. Furthermore in the limit $\L \to \infty$ we have that $A \to I$.
 Taking two functional derivatives of $A$  we see that
\beq
A_{;ab} = \Delta_{ac} \left(c^{-1}(\Delta)\right) \Pi^{c}\,_b  + \mathcal{O}(\E)
\eeq
and thus at the level of the the two point function $c$ is a high momentum cutoff function for the effective propagator $c(\Delta) \Delta^{-1} \gamma^{-1}$ on the transverse space which reduces to the propagator for the action $I$ at low momentum since $c(0) = 1$. For our example of Einstein gravity $A$ is given by
\beq
A=    \frac{Z}{16 \pi G_N} \int d^Dx \sqrt{\det g}  \left[( -R + 2 \bar{\lambda} ) + \L^{-2} (G^{\mu\nu} + \bar{\lambda} g^{\mu\nu}) X(\Delta) \left(R_{\mu\nu}  - \frac{2 \bar{\lambda} }{D-2}  g_{\mu\nu} \right)  \right]
\eeq
where here $G^{\mu\nu}$ and $R_{\mu\nu}$ are the Einstein tensor and Ricci tensors.

Due to the fact that $G$ is bilinear in $C$ we need also a bi-linear structure for $S^{(2)}$. Following \cite{Bakeyev:1996is} this is achieved by including a term in the action which is bi-linear in $A_{,a}$.
Here we will therefore choose $S_{\rm UV} =   \frac{1}{2} A_{,a} \gamma^{ab} A_{,b}$ such that 
\beq \label{S_form}
S = I  + \S   +   \frac{1}{2} A_{,a} \gamma^{ab} A_{,b} \,,
\eeq
where $\S$ denotes other possible `interaction' terms that can also be present in the action.
The two point function will therefore diverge as $S_{;ab} \sim  A_{;ca} \gamma^{cd} A_{;db}$.
To fully regularise the theory we demand that \eq{G_condition}, \eq{B_condition} and \eq{b_condition} are satisfied  which is achieved by setting
\bea \label{PVs}
( C^{-1})^a\,_b = \delta^a\,_b + \gamma^{ac} A_{;cb}   + K^a_{\a} Y^{\a\b} K_{\b b }   \,, \nn
B^{\a}\,_\b =  \delta^\a_\b  +  Y^{\a\c} \Delta^\parallel_{\c\b} \,,\\
b^\a\,_\b = \mathfrak{c}^\a\,_\b + (\Delta^\parallel)^\a\,_\b\,, \nonumber
\eea 
where  the operators $Y^\a\,_\b = (Y(\Delta^\parallel))^\a\,_\b $ and $   \mathfrak{c}^\a\,_\b = \mathfrak{c}(\Delta^\parallel)$ are functions  of $\Delta^\parallel$.
Here $Y(z)$ should diverge for large momentum ( it is sufficient that $Y(z)$ diverges at the same rate as $c^{-1}(z)$) and go to a constant or vanish at low momentum while $ \mathfrak{c}(z)$ should vanish for large momentum and for low momentum $ \mathfrak{c}(0) =1$.

 Let us now see that all the conditions are met. First we insert the expression \eq{PVs} for $C^{-1}$ into \eq{G} and then apply the transverse projection operators to obtain  
 \bea
\Pi^c\,_a G_{cd} \Pi^d\,_b &\sim&  \Pi^{c}_{a} A_{;c}\,^e A_{;ed} \Pi^d\,_b \nn 
&\sim&  A_{;a}\,^e A_{;eb}  \sim  S_{;ab}
  \eea
 where to arrive at the second line we used that $A_{;ab} \Pi^b_c = A_{;ac} + A_{,b} K^b_{\a; a} \mathfrak{G}^{\a\b} K_{\b c} \sim A_{ac}$. Thus we see that \eq{G_condition} is satisfied. Then we insert the expression for $C^{-1}$ into  \eq{B_condition} 
\bea
C^{a}_b K^b_\gamma &=&  K^a_\gamma- \gamma^{ac} A_{,b}   K^b_{\gamma;c}  + K^a_{\a} Y^{\a\b} K_{\b b } K^b_\gamma \nn
& \sim&   K^a_{\a} Y^{\a\b} K_{\b b } K^b_\gamma\nn
& \sim&  K^a_\a B^{\a}\,_\c \,.
\eea
Lastly \eq{b_condition} follows since at high momentum $\mathfrak{c} \sim 0 $.

\subsection{Discussion and summary}
Let us comment that although the regularisation scheme introduces some non-localities these are all quasi-local meaning one can still perform a derivative expansion and hence the regularisation does not introduce spurious IR singularities. Indeed, due to the fact that gauge transformations do not leave momentum eigenstates invariant, it hard to imagine that any gauge invariant regularisation can be achieved without the introduction of some sort of non-locality. 
A particular quasi-locality can be seen if we include the PV operators as terms in the action. For example we can write
\beq
(\Det C)^{-1} = e^{- \Tr \log C }
\eeq
which can be evaluated using heat kernel techniques to yield non-local terms.
However, since our case the PV operators are massive   $\Tr \log C$ will be quasi-local. Additionally taking the limit $\L \to \infty$ the non-localities are absent since $C$, $B$ and $b$ are just the identity operator and thus we recover strict locality in the continuum limit provided that $\S$ is also local in this limit.

In the limit $\Lambda \to 0$ we will lose quasi-locality since a derivative expansion is not possible. However since we have regulated all loops at the scale $\L$ taking $\L \to 0$ must set all loop corrections to zero and the semi-classical approximation to the path integral becomes exact. Consequently, if we renormalise the theory such that $\mathcal{Z}$ is independent of $\L$ the path integral is then given by 
\beq \label{ZL0}
\mathcal{Z} =\sum e^{-S[\bar{\phi}]} |_{\L = 0}
\eeq
where we should sum over instantons.

To summarise the path integral \eq{ZGaugeInv} is gauge invariant and thus formally independent of how we perform the integration over the space of orbits and can be written as  \eq{pathintegral}. The action is given by \eq{S_form} with the metrics given explicitly by \eq{G} and \eq{H} with \eq{PVs}. These choices ensure that momentum integrals are cutoff in the UV by the scale $\L$. This regularisation works without introducing any background field and independent of any gauge fixing condition.
We have also written the path integral in a manifestly field covariant manner such that the choice of field variables does not affect the physics. It should be noted that the choice of the measure does affect the physics and thus the choice of $G$ and $H$ can in principle lead to different predictions \cite{Falls:2018olk}. Since we will demand that the physics only depends on the continuum limit the choices of  
$\gamma$ and $\eta$ are physical.

\section{Flow equation}
\label{secIII}

We now want to write down a renormalisation group flow equation for $S$  based on the regularisation scheme introduced in the last section. 
Ultimately this relates the regularisation scheme to a particular averaging of the bare fields $\varphi$ which we understand as a continuum version of Kadanoff blocking \cite{Kadanoff:1966wm}.
Here we shall require the flow equation for $S$ to be a suitable generalisation of the Wilson-Polchinski equation \cite{Wilson:1973jj,Polchinski:1983gv} satisfying the following set of criteria:
\begin{enumerate}[label=(\roman*)]
\item  {\it a solution to the flow equation yields a regularised path integral with momentum cutoff at the scale $\L$, }  
\item  {\it the path integral $\mathcal{Z}$ is  independent of $\L$}, 
\item  {\it the description of the physics is unchanged at low momentum scales compared to $\L$, }
\item {\it the flow equation for $e^{-S}$ can be written as a generalised heat equation of the form
\beq
(\L \partial_\L +\Op_\L )    \frac{\sqrt{\det G[\phi]  } }{\sqrt {\det H[\phi]} }  e^{-S[\phi]}  = 0
\eeq
where $ \Op_\L$ is a second order functional differential operator.}
 \end{enumerate}
  Together (i), and (ii) mean that as we decrease $\L$ we are indeed integrating out the high momentum modes. This follows since (i) implies that the modes with $p^2\gg \L^2$ are always suppressed while (ii) makes sure these modes are being integrated into $S$. The requirement (iii) is needed to ensure that we are only integrating out the high energy modes leaving modes $p^2 \ll \L$ unintegrated. In practice this will be satisfied provided we can expand the flow equation in a derivative expansion which will be the case if all the ingredients are quasi-local. In real space this means that we are averaging fields only on local patches of the manifold.   
We impose criteria (iv) such that the flow equation for $S$ is non-linear  which is necessary to describe perturbations of renormalisation group fixed points \cite{Wegner1974}.

General schemes which satisfy the requirements (i)-(iv) have been considered previously for  scalar fields \cite{Arnone:2002yh,Arnone:2003pa} 
as well as for gauge theories \cite{Arnone:2002qi, Arnone:2002fa, Arnone:2002cs}. There it has also proven useful to fix the classical two point function in order to satisfy (i). Here we will similarly require that
\beq \label{Classical_action}
S_0 =  I  +  \frac{1}{2} A_{,a} \gamma^{ab} A_{,b}  +  \S_0  \,,  \,\,\,\,\,\,\,\,  {\rm where} \,\,\,  \S_0 = O(\E^3)\,.
\eeq 
is a solution to the classical flow equation (obtained by setting $S \to S/\hbar$ and taking the limit $\hbar \to 0$).
Since $\S_0$  is third order in the equations of motion $\E$ the two point function is fixed to given by $S_{0;ab}  = I_{;ab}  + A_{ca} \gamma^{cd} A_{;db}$ for any $\phi$ which satisfies $\E[\phi] =0$.

\subsection{General structure}
\label{Gen_Stru}

As first discussed in  \cite{Wegner1974}, a large number of flow equations which satisfy (ii) can be parameterised by the choice of a field redefinition since if a change in $\L$ can be expressed as a field redefinition the path integral will be invariant. This understanding also explains why physics is independent of the scheme since it follows that different schemes are also related by field redefinitions \cite{Latorre:2000qc}.
Geometrically we can express a field redefinition as a diffeomorphism generated by a vector $\Psi^a$ on $\Phi$ which induces a diffeomorphism on $\Phi/\G$ which keeps \eq{pathintegral} invariant.  We can therefore write flow equations concisely as
\beq \label{flow_general}
\L D_\L      \frac{\sqrt{\det G[\phi]  } }{\sqrt {\det H[\phi]} }  e^{-S[\phi]} = 0
\eeq
 where $\L D_\L$ denotes the `total renormalisation group derivative' 
\beq
\L D_\L \equiv  \dL  + \mathcal{L}_{\Psi}
\eeq
given in terms of the Lie derivative   $\mathcal{L}_{\Psi}$.
 The $\L D_\L$ derivative of $S$, $G$ and $H$  are given by\footnote{This follows since $G_{ab}$ is a tensor on $\Phi$ and $H_{\a\b}$ is  a scalar on $\Phi$.} 
\bea
\L D_\L S  &\equiv&  \dL S  +     \Psi^a S_{,a} \,,\\[5pt]
\L D_\L G_{ab}  &\equiv&  \dL G_{ab}  +     \Psi^c G_{ab,c}  +  \Psi^c\,_{,a} G_{bc} +   \Psi^c\,_{,b} G_{ac}\,,\\[5pt]
\L D_\L H_{\a\b} &\equiv&  \dL  H_{\a\b}  +     \Psi^c  H_{\a\b,c}  \,.
\eea
  We can then equivalently write \eq{flow_general} as
\beq \label{The_Flow_Equation}
\L D_\L S =    \hbar   \frac{1}{2} \Tr[ G^{-1} \L D_\L G - H^{-1} \L D_\L H]
\eeq
where we have made explicit the dependence on $\hbar$. 
The trace $\Tr[...]$ means that we take the traces of both the operators inside the square brackets. The PV regularisation involves a cancelation between the two terms and hence the trace must be taken by combining traces. The way in which this calculation occurs will be described in section~\ref{cutoff_sec} allowing us to write the trace as two separate traces which are separately regularised. 
The operator under the trace is also quasi-local provided $\Psi^a$ is quasi-local ensuring that the trace is regularised in the IR.
We stress that the flow equation \eq{The_Flow_Equation} is obtained without reference to any gauge fixing procedure.
In appendix~\ref{GFflow} we show that the same flow equation can be derived from the gauge fixed path integral independently of the gauge condition.

A perturbative expansion around a Gaussian fixed point can be achieved by expanding 
\beq \label{Pert_Ex}
S = S_0 + \hbar S_1 + ...
\eeq
where $S_0$ is the `classical action' and $S_\ell$ for $\ell>0$ are the $\ell$-loop corrections.
The classical action then obeys the classical flow equation
\beq
\L D_\L S_0 = 0 
\eeq
which we require to have a solution of the form \eq{Classical_action}.

Here we take $\Psi^a$ to have the form
\beq\label{Psi_0}
\Psi^a =  - \frac{1}{2} \K^{ab} \Sigma_{,b} + \psi^a  \,,
\eeq
where $\K^{ab}$ and $\psi^b$ are independent of $S$ and we define the functionals
\bea \label{Sigma}
\Sigma &=&  S - \hat{S} \, \\
\label{Shat}
\hat{S} &=& A + \frac{1}{2} A_{,a} \gamma^{ab}   A_{,b}
\eea
where it follows, from comparison with \eq{S_form}, that $\Sigma \equiv \S  - \frac{1}{2} I_{,a} X^{ab} I_{,b}$. 
Since $\Sigma$ is linear in $S$ this ensures that we satisfy (iv) with
\beq
\Op_\L  = \frac{\delta}{\delta \phi^a}   \left[ \frac{1}{2} \K^{ab}  \left(   \frac{\sqrt{\det G[\phi]  } }{\sqrt {\det H[\phi]} } \frac{\delta}{\delta \phi^b}  \frac{\sqrt {\det H[\phi]}}{\sqrt{\det G[\phi]  } }    + \hat{S}_{,b} \right)  +  \psi^a  \right]\,,
\eeq
where the functional derivatives act on everything to the right.

\subsection{ERG kernels}
\label{kernels}
It remains to fix the form of $\K^{ab}$ and $\psi^b$ which we refer to as the {\it ERG kernels}. To adhere to the requirement (iii) we must make sure both $\K$ and $\psi$ are quasi-local. 
The greatest restriction on $\K^{ab}$ and $\psi^b$ comes from requiring  (i) to hold. Practically this means that the ERG kernels must be related to the regularised action ensuring that solutions to the flow for $\S$ do not generically destroy the regularisation scheme.
This will be satisfied  provided the $n$-point $\S^{(n)}$ functions of $\S$ do not diverge too strongly relative to the UV cutoff.  In particular we will suppose that $\S^{(n)}$ diverges maximally as
\beq \label{Max_Div}
\S^{(n)} \sim A^{(n)}
\eeq
then we need to make sure that no terms are present in the flow equations for $\S^{(n)}$ that diverge faster than \eq{Max_Div}.
This being the case the cutoffs can be made strong enough to regularise the theory e.g. by choosing $c(z) = e^{-z}$. 

An important property of $\hat{S}$ given by \eq{Shat} is that its first derivative can be written as
\beq \label{ShatEofM}
\hat{S}_{,b} = A_{,a} (\delta^a\,_b + \gamma^{ac} A_{;cb} ) = A_{,a} ( C^{-1})^a\,_b 
\eeq
where the second equality follows from $K^a_\a A_{,a} = 0$.
Using the identity \eq{ShatEofM} the LHS of the flow equation can be written as
\bea \label{flow_LHS}
\L D_\L S  &=&  \dot{A}_{,a} \gamma^{ab} A_{,b}   +     \frac{1}{2} A_{,a} \dot{\gamma}^{ab} A_{,b}     +   A_{,a} ( C^{-1})^a\,_b   \psi^b   - \frac{1}{2}   A_{,c} ( C^{-1})^c\,_a    \K^{ab} \Sigma_{,b} \nn
 &&  \,\,\,\,\, \,\,\,\,\, \,\,\,\,\,  \,\,\,\,+   \dL A   +  \L \partial_\L \Sigma  - \frac{1}{2}  \Sigma_{,a}  \K^{ab} \Sigma_{,b}   +    \Sigma_{,a} \psi^a 
\eea
where the ``  $\dot{}$  '' denotes a $\dL$ derivative.
The terms appearing on the first line of \eq{flow_LHS} are bilinear in $A_{,a}$ and $A_{;ab}$. As a result they can potentially destroy the regularisation since they can contribute terms to the flow of $\S$  which diverge faster than \eq{Max_Div}. We must therefore choose $\psi^a$ and $\K^{ab}$ to cancel the offending terms. First let us deal with the first three terms on the RHS of \eq{flow_LHS} which do not involve $\K$. These can be canceled identically by choosing $\psi$ to be given by
\beq \label{psi_0}
\psi^c=-  C^{c}\,_b  \left(  \gamma^{ba}   \dot{A}_{,a}   +   \dot{\gamma}^{ba}  \frac{1}{2} A_{,a}     \right) \,.
\eeq
This leaves the  fourth term in  \eq{flow_LHS} which must be rendered harmless by choosing $\K^{ab}$ appropriately.
 Let's dissect the fourth term a little by taking two functional derivatives.  Then we will obtain terms such as 
 \beq
 \dL \S_{;de} = \frac{1}{2}   A_{,ce}  \gamma^{cf}  A_{;fad}    \K^{ab} \Sigma_{,b} + ...
 \eeq
 which contributes to the flow of $\S_{,de}$ which diverges with order $\sim A^{(2)}A^{(3)}  $. In order to remove such divergences we take
  \beq \label{K_kernel}   
\K^{ab} = C^{a}\,_c \kappa^{cb}  \,,
\eeq  
such that we cancel the factor of $C^{-1}$ and we take  $\kappa^{a}\,_b = (\kappa(\Delta))^{a}\,_b$ to be a function of $\Delta$. 
Finally we fix $\kappa^{ab}$ by our requirement that we have a classical solution of the form \eq{Classical_action}.
 The classical flow is given by
\beq \label{tree_level_flow}
\L \partial_\L \Sigma + \L \partial_\L A - \frac{1}{2} ( \Sigma_{,a} +   \hat{S}_{,a} )  C^{a}\,_c \kappa^{cb}  \Sigma_{,b}  -   \Sigma_{,c} C^{c}\,_b  \left(  \gamma^{ba}   \dot{A}_{,a}   +   \dot{\gamma}^{ba}  \frac{1}{2} A_{,a}     \right)    = 0 + \mathcal{O}(\hbar) \,.
\eeq
now the task is to insert \eq{Classical_action} into \eq{tree_level_flow} dropping all terms which are order $\E^3$ and solve for $\kappa^{ab}$.
We carry out this short calculation in Appendix  \ref{kappa_details} arriving at the explicit form
\bea  \label{kappa}
\kappa^{ab} 
=\left(\frac{-4 \Delta c'(\Delta) + c(\Delta) (2 + \upeta)}{ c^2(\Delta)  + \Delta}  \right)\,^a\,_c  \gamma^{cb} \,.
\eea
It is  evident from the form of $\kappa$ that it behaves as a UV cutoff.

Now we consider the form of the trace on the RHS of \eq{The_Flow_Equation} since this will contribute to the flow of $\S$.  The terms in the operator under the trace can be written in terms  of $C$, $B$ and $b$ which do not have the dangerous UV behaviour of $\hat{S}$. For example
\beq
(G^{-1})^{ab} \L D_\L  G_{ab} =  C^{ab}   \L D_\L (C^{-1})_{ab}   +  C^{a}\,_b   \L D_\L (C^{-1})^b\,_a 
\eeq
thus in expanding the trace it is evident that we do not obtain terms in the flow for $\S$ that destroy the regularisation. 
This is how the resolution of the problem of overlapping divergencies manifests at the level of the flow.

\subsection{UV cutoff in the flow}
\label{cutoff_sec}

Since the path integral is regulated in the UV we expect that we should have no UV divergences in the flow equation.
This means that the operator under the trace in the RHS of \eq{The_Flow_Equation} should go to zero in the high momentum limit $p^2/\L^2 \to \infty$ which occurs when the eigenvalues of $\Delta$ and $\Delta^\parallel$ diverge. 
Let us now demonstrate that this will be the case.
First we note that for high momentum 
\beq \label{hatP}
 G_{ab} \sim  \hat{S}_{;ab}   +  K_{a \a} Y^{\a\b} \Delta^\parallel_{\b \c} Y^{\c\de}  K_{b \de} \equiv \hat{P}^{-1}_{ab}\,,
\eeq
while 
\beq
H_{\a\b} \sim
( \Delta^\parallel Y \Delta^\parallel Y \Delta^\parallel)_{\a\b} \,.
\eeq
The terms in $\L D_\L G$ and $\L D_\L H$ where the $\L D_\L$ acts on the sub-leading terms of $G$ and $H$ are less divergent than $G$ and $H$ so the corresponding terms in   \eq{The_Flow_Equation}  are cutoff by the factors $G^{-1}$ and $H^{-1}$.  Therefore the potentially divergent part of the trace is given by
\beq \label{trace_div}
\Tr [G^{-1} \L D_\L \hat{S}^{(2)}  + G^{-1} \L D_\L  KY \Delta^\parallel Y K   - H^{-1}  \L D_\L \Delta^\parallel Y \Delta^\parallel Y \Delta^\parallel]\,.
\eeq
 To show that \eq{trace_div} will also be UV regulated relies on the properties of the ERG kernels $\psi$ and $\K$ as well as the relation between $G$ snd $H$.
First let us note that we have the two identities
\beq \label{Shatflow}
 \dL \hat{S}  +  \psi^a \hat{S}_{,a} = \L \partial_\L A
\eeq
and
 \beq \label{KShat}
\hat{S}_{,a} \K^{ab}  =   A_{,a}  \kappa^{ab}  \,,
\eeq
which relate $ \hat{S}$ to the RG kernels.
If we differentiate these identities we obtain terms which appear in   $\L D_\L \hat{S}_{,ab}$.
Explicitly differentiating \eq{Shatflow} twice we have
\beq
(\dL + \mathcal{L}_\psi) \hat{S}_{;ab} = - \psi^c_{;ab} \hat{S}_{,c}  +  \L \partial_\L A_{;ab}  + \psi^c R^d\,_{abc} \hat{S}_{,d}\,,
\eeq
where $\mathcal{L}_\psi$ is the Lie derivative with respect to $\psi$ and $R^d\,_{abc}$ is the Riemann curvature for the metric $\gamma_{ab}$.
Furthermore differentiating \eq{KShat} twice we obtain
\beq
 \hat{S}_{;cb}  \K^{cd}_{;a} +  \hat{S}_{;ca} \K^{cd}_{;b}  +  \hat{S}_{,c}    \K^{cd}_{;ab} +   \hat{S}_{;abc}  \K^{cd}  =  \nabla_b \nabla_a  (A_{,c}  \kappa^{cd})  + R^e\,_{abc} \hat{S}_{,e}    \K^{cd}  \,.
 \eeq
 Using these two identities we find that the operator $\L D_\L \hat{S}_{;ab} $ is given by
\bea
\L D_\L \hat{S}_{;ab} &=&  - \psi^c_{;ab} \hat{S}_{,c}  +  \dL A_{;ab}  - \frac{1}{2} \left(  \hat{S}_{;bc}   \K^{cd} \Sigma_{;da}  + \hat{S}_{;ac}\K^{cd} \Sigma_{;db}  -  \hat{S}_{,d}    \K^{dc}_{;ab}  \Sigma_{,c} +   \Sigma_{,c}  \nabla_b \nabla_a   A_{,d}  \kappa^{dc}  \right)  \nn
&&+ \Psi^c R^d\,_{abc} \hat{S}_{,d}\,.
\eea
One can then check that each term diverges less quickly than $G$ and therefore  $G^{-1} \L D_\L \hat{S}^{(2)}$ will decrease rapidly for high momentum provided the regulator $c$ is strong enough. This leaves the potentially divergent trace
\beq \label{trace_div_K}
\Tr [G^{-1} \L D_\L \gamma K Y  \Delta^\parallel Y K \gamma  - H^{-1} \L D_\L  \Delta^\parallel Y \Delta^\parallel Y \Delta^{\parallel}]
\eeq
which relies on the relation between $G$ and $H$ to be regularised.
In particular we can rewrite this trace as
\beq \label{mixed_trace}
\Tr [G^{-1}  \L D_\L \gamma K Y  \Delta^{\parallel} Y K \gamma  - H^{-1} \L D_\L  \Delta^{\parallel} Y \Delta^{\parallel}Y \Delta^{\parallel}]  = \Tr [(G^{-1} -  K H^{-1} K )  \L D_\L \gamma K Y \Delta^{\parallel} Y K \gamma ]
\eeq
where we have used that $[\L D_\L,K] = 0$ and the cyclic nature of the trace. 
Then we can use that at high momentum  $G^{-1}$ can be expressed as 
\beq
G^{-1} \sim \hat{P} 
\eeq
where $\hat{P}$ is the propagator corresponding to the operator defined in \eq{hatP}.
Since $\hat{S}_{;ab} K^b_{\a} =- \hat{S}_{,b} K^b_{\a;a}$ the propagator $\hat{P}$ can be decomposed into transverse and longitudinal parts in the high momentum limit
\beq
\hat{P} \sim   \Pi \hat{P}^{\bot} \Pi + K P^{\parallel}K
\eeq
with the transverse propagator satisfying
\beq
\Pi \hat{S}^{(2)}\Pi    \hat{P}^{\bot} = \Pi 
\eeq
and the longitudinal propagator defined by
\beq \label{Ppara}
 P^{\parallel} =  \frac{1}{\Delta^{\parallel} Y  \Delta^{\parallel}    Y \Delta^{\parallel}}   \sim  H^{-1}   \,.
\eeq
It then follows that the trace \eq{mixed_trace} is cutoff in the UV since the terms of order  $P^{\parallel}$ cancel and the term of order $P^{\bot}$ also vanishes by 
 \beq \label{Long_DgKYKg}
 \Pi  ( \L D_\L \gamma K Y \Delta^{\parallel} Y K \gamma)  \Pi = 0
 \eeq
 where $\L D_\L$ only acts on the the expression in the brackets.
 
  Thus we can conclude that the momentum integrals in the trace will be cutoff at $\L$. In particular the trace can be split into the following two traces
  \beq
  \frac{1}{2} \Tr[ G^{-1} \L D_\L G - H^{-1} \L D_\L H] =  \mathcal{T}_\Phi +  \mathcal{T}_\G 
 \eeq
 where
 \beq
 \mathcal{T}_\Phi = \frac{1}{2} \Tr[ G^{-1} \L D_\L  G  -  K H^{-1} K   \L D_\L \gamma K Y \Delta^{\parallel} Y K \gamma ] 
 \eeq
 and
\beq
\mathcal{T}_\G = -  \frac{1}{2}   \Tr[ H^{-1} \L D_\L (H- \Delta^\parallel Y \Delta^\parallel Y \Delta^\parallel ) ]
\eeq
which are separately UV regularised.

\section{One-loop renormalisation}
\label{secIV}
Let us now demonstrate how the flow equation can be used to compute the one-loop renormalisation of the Wilsonian effective action.

\subsection{Scheme independence}
\label{Scheme_Ind}
Before obtaining  explicit beta functions it is informative to show how scheme independent results can be computed by making as few assumptions as possible. Let us therefore now keep $\Psi^a$ quite  general and aim to show how the standard one-loop renormalisation can be extracted. We will assume that $\Psi^a$ is gauge covariant and that the solution to the flow equation leads to a regularised path integral. Since the scheme independent renormalisation corresponds to the renormalisation of terms which do not vanish on the equations of motion we will expand around a solution to the equation of motion for $S[\phi]$.

To perform the loop expansion we insert \eq{Pert_Ex} into the the flow equation. The flow of the classical action $S_0$ takes the form
\beq \label{S0flow}
 \dL S_0  + S_{0,a} \Psi^a = 0\,.
\eeq
We now expand the flow for $S_0$ around an on-shell field configuration $\phi = \bar{\phi}$ which satisfies $S_{,a}[\bar{\phi}] = 0$. In general we note that $\bar{\phi}$ can depend on $\L$ since even at the classical level the flow equation is non-trivial even though it does not contain any loop integrals. The expansion is given by
\beq
S[\bar{\phi}] + \frac{1}{2} S_{,ab}[\bar{\phi}]  (\phi^a - \bar{\phi}^a)  (\phi^b - \bar{\phi}^b) +...
\eeq
 The first order in the on-shell expansion consists of simply setting $\phi = \bar{\phi}$ in \eq{S0flow}  to obtain 
\beq
\dL S_0[\bar{\phi}] = 0 \,.
\eeq
Since we are on shell we also have that the total derivative vanishes
\beq \label{S0_L_indep}
 \frac{d}{d\L}  S_0[\bar{\phi}]  =  \dL S_0[\bar{\phi}] + \dL S_{0,a}[\bar{\phi}] \dL \bar{\phi}^a = 0 
\eeq
 Next we differentiate \eq{S0flow} once and then setting $\phi = \bar{\phi}$ we obtain
\beq \label{S0flow1}
\dL S_{0,a}[\bar{\phi}]  + S_{0,ba}[\bar{\phi}] \Psi^b[\bar{\phi}] = 0\,,
\eeq
since $S_{0,a}[\bar{\phi}]  = 0$  we also have that
\beq 
 \frac{d}{d\L} S_{0,a}[\bar{\phi}] = \dL S_{0,a}[\bar{\phi}]   + S_{0,ba} \dL \bar{\phi}^b = 0
\eeq
and thus we can conclude that 
\beq
 S_{0,ba}[\bar{\phi}] (\Psi^b[\bar{\phi}] -  \dL \bar{\phi}^b) =0 \,.
\eeq
For a non-gauge theory we could then infer that $\Psi^b[\bar{\phi}] =  \dL \bar{\phi}^b$.
However, since for a gauge theory $S_{0,ba}[\bar{\phi}] $ is not invertible we can only conclude that the weaker identity 
\beq \label{dLbarphi}
 \Pi^{a}\,_b[\bar{\phi}]   \Psi^b[\bar{\phi}]   =   \Pi^{a}\,_b[\bar{\phi}]  \dL \bar{\phi}^b
\eeq
holds in this case.
Differentiating \eq{S0flow} twice and going on-shell we then have the identity
\beq \label{S0flow2}
 \dL S_{0,ab}[\bar{\phi}]   +  S_{0,c ab}[\bar{\phi}]  \Psi^c[\bar{\phi}]  + S_{0,ca}[\bar{\phi}]  \Psi^c\,_{,b}[\bar{\phi}]  +  S_{0,cb}[\bar{\phi}]  \Psi^c\,_{,a}[\bar{\phi}]   = 0\,
\eeq
which gives the flow of on-shell two point function $S_{0,ab}$. Although we have not solved the flow for $S_0[\phi]$ completely the current order is enough to obtain the flow of the one-loop effective action
$S_1[\phi]$ evaluated at $\phi = \bar{\phi}$.
 From the flow of $S_1[\phi]$ we have that 
\beq \label{S1flow0}
\L \partial_\L S_1[\bar{\phi}]  +     S_{1,a}[\bar{\phi}] \Psi^a[\bar{\phi}]  = \frac{1}{2} \Tr [ G^{-1}[\bar{\phi}]  \L D_\L G[\bar{\phi}]  - H^{-1}[\bar{\phi}] \L  D_\L H[\bar{\phi}] ]|_{S = S_0}  \,.
\eeq 
Since $K^a_{\a} S_{1,a} = 0$, and thus $  S_{1,a} = \Pi^b\,_{a} S_{1,b}$, we can insert $\Pi$ into the LHS of \eq{S1flow0}  and use \eq{dLbarphi} to arrive at
\beq \label{total_derivative}
\L \partial_\L S_1[\bar{\phi}]  +     S_{1,a}[\bar{\phi}] \Psi^a[\bar{\phi}]  = \L \frac{d}{d\L} S_1[\bar{\phi}] 
\eeq
and thus the LHS of \eq{S1flow0} can be written as a total derivative
\beq \label{one_loop0}
\frac{d}{d\L} S_1[\bar{\phi}]  = \frac{1}{2} \Tr [ G^{-1}[\bar{\phi}] \L D_\L G[\bar{\phi}]  - H^{-1}[\bar{\phi}]  \L D_\L H[\bar{\phi}] ]  |_{S = S_0} \,.
\eeq

The expression \eq{one_loop0} has been found using the flow equation where no gauge fixing has taken place it also appears to depend on the specific choice for $\Psi$. 
Let us now show that the RHS of \eq{one_loop0} can be brought into the form obtained by instead carrying out the calculating the gauge-fixed functional integral and evaluating the one-loop determinants in the usual manner without reference to the form of $\Psi$. To this end we define the propagator $P$ by
\beq \label{P}
P^{-1}_{ab} = S_{,ab}  + K_{a\a} Y^{\a\b}  \Delta^{\parallel}_{\b\c}   Y^{\c\de} K_{\de b} 
\eeq
where here, and in the rest of this subsection, we take $\phi = \bar{\phi}$.
The propagator $P$ for \eq{P} is in a background field gauge $F_{\a} = K_{\a a} \phi^a$ and where the gauge fixing action is chosen to include higher derivatives with  $S_{\rm gf} = \frac{1}{2}  F_\a Y^{\a\b}   \Delta^{\parallel}_{\b\c}   Y^{\c\de} F_{\de}$ with the higher derivatives entering via the choice of averaging the gauge condition.   The corresponding Faddeev-Popov and the third ghost determinants can be written as
\beq
\det [\Delta^{\parallel}_{\a\b} ] \sqrt{\det [Y^{\a\b}    \Delta^{\parallel}_{\b\c}    Y^{\c\de}]}  =   \sqrt{\det[ P_{\parallel}^{-1}  ]} \,,
\eeq
where  $P_{\parallel}$ is given by \eq{Ppara}.
Note that since we are on-shell $S_{,ab} K^b_\a =0$ and therefore $P$ can be split into longitudinal and transverse terms
\beq
P = \Pi P_\perp \Pi + K  P_{\parallel} K
\eeq
where $S^{(2)} P_\perp = \Pi$.
To one-loop order we have that the path integral is given by  
\beq
\mathcal{Z} = e^{- \frac{1}{\hbar} S_0 - S_1} \frac{\sqrt{\det[ G P]}}{\sqrt{\det [H  P_{\parallel}]}} \,,
\eeq
where we have made the saddle point approximation and performed the one-loop gaussian integral.
We then note that the one-loop determinants  $\det [H  P_{\parallel}]$ and $\det[ G P]$ are both UV finite.   
Demanding that the one-loop path integral is independent of $\L$ then leads to the expression
\beq \label{one_loop}
\frac{d}{d\L}   S_1 =  \frac{1}{2}  \frac{d}{d\L}  \Tr \log[P G] -  \frac{1}{2} \frac{d}{d\L}   \Tr \log[ P_{\parallel} H]  
\eeq
where we have used \eq{S0_L_indep}.
It then remains to demonstrate that \eq{one_loop} is reproduced by the flow equation \eq{one_loop0}.
To show this we first note that since $P$, $G$, $Q$ and $H$ are covariant we have that $ \Tr \log[PG] $ and $\Tr  \log[Q H] $ are both gauge invariant and therefore  (by the same argument that lead to \eq{total_derivative}) we have
\bea  \label{TotalG}
&&\frac{d}{d\L}    \Tr \log[P G] =     \left(\dL  + \Psi^{c} \frac{\delta}{\delta \phi^c}\right)   \Tr \log[P G]   \\
&&= (G^{-1})^{ab}(   \dL G_{ab}  + G_{ab,c} \Psi^c)   -   P^{ab}  ( \dL   S_{,ab}   +  S_{,abc} \Psi^c    +    \left(\dL  + \Psi^{c} \frac{\delta}{\delta \phi^c}\right)  K_{a\a} Y^{\a\b} \Delta^{\parallel}_{\b\c}   Y^{\c\de} K_{\de b}) \nonumber
\eea
and
\bea \label{TotalH}
\frac{d}{d\L} \Tr  \log[ P^{\parallel} H]   &=&  \left(\dL  + \Psi^{c} \frac{\delta}{\delta \phi^c}\right)   \Tr \log[P^{\parallel} H]   \\
&=&   \Tr[ H^{-1} \L D_\L H]  -   P_{\parallel}^{\a\b}   \left(\dL  + \Psi^{c} \frac{\delta}{\delta \phi^c}\right)  ( P_{\parallel}^{-1})_{\a\b} \nonumber
\eea
Then  we can use \eq{S0flow2} and the fact that $P$ is a propagator for $S_{,ab}$ in the transverse space, meaning that $P^{ab} S_{,bc} = \Pi^a\,_c$, to express \eq{TotalG} as
\beq
\frac{d}{d\L}    \Tr \log[P G]  =   (G^{-1})^{ab}(   \dL G_{ab}  + G_{ab,c} \Psi^c)   + 2   \Pi^a\,_b \Psi^b\,_{,a}    -   P^{ab}   \left(\dL  + \Psi^{c} \frac{\delta}{\delta \phi^c}\right)  K_{a\a} Y^{\a\b} \Delta^{\parallel}_{\b\c}   Y^{\c\de} K_{\de b}\nn
\eeq
inserting $\Pi = \delta - \Xi$, using the definition of $\L D_\L G_{ab}$ and noting that $P^{ab}$ can be replaced by the longitudinal part $K P_{\parallel} K$ in the last term we then have
\beq 
\frac{d}{d\L}    \Tr \log[P G] =     (G^{-1})^{ab}(  D_\L G_{ab})  -   2   \Xi^a\,_b \Psi^b\,_{,a  }  -  K^b_\de P_{\parallel}^{\de\c} K^a_\c   \left(\dL  + \Psi^{c} \frac{\delta}{\delta \phi^c}\right)  K_{a\a} Y^{\a\b} \Delta^{\parallel}_{\b\c}   Y^{\c\de} K_{\de b} \,.
\eeq
Now we note that
\beq
    P_{\parallel}^{\a\b}   \left(\dL  + \Psi^{c} \frac{\delta}{\delta \phi^c}\right)  (P_{\parallel}^{-1})_{\a\b} =        2   \Xi^a\,_b \Psi^b\,_{,a  }  +   K^b_\de  P_{\parallel}^{\de\c} K^a_\c   \left(\dL  + \Psi^{c} \frac{\delta}{\delta \phi^c}\right)  K_{a\a} Y^{\a\b}\Delta^{\parallel}_{\b\c}   Y^{\c\de} K_{\de b}
\eeq
where the term  $ 2   \Xi^a\,_b \Psi^b\,_{,a  }$ comes from the $\phi$-derivative hitting the $K$s at the far left and far right of $ P_{\parallel}^{-1}$ and using the Ward identity  $\mathcal{L}_{K_\a}\Psi^b = 0$.
So finally we have
\bea \label{TotalG}
\frac{d}{d\L}    \Tr \log[P G] 
&=&   \left(   (G^{-1})^{ab} (  D_\L G_{ab})   -   P_{\parallel}^{\a\b}  \left(\dL  + \Psi^{c} \frac{\delta}{\delta \phi^c}\right)  (  P_{\parallel}^{-1})_{\a\b}  \right) \,.
\eea
Using \eq{TotalH} and \eq{TotalG} we see that the RHS of  \eq{one_loop} is given by the RHS of \eq{one_loop0}. Thus the flow equation, which was derived without fixing the gauge, reproduces the standard result obtained by gauge fixing. Furthermore the gauge fixed expression is independent of $\Psi^a$ indicting the scheme independence.

\subsection{Explicit evaluation of the one-loop trace}
\label{one_loop_trace}

Let us now work with the explicit form of $\Psi^a$ given by \eq{Psi_0} with \eq{psi_0}, \eq{K_kernel} and \eq{kappa} such that 
explicitly we have
\beq
\Psi^a =  - \frac{1}{2}  C^{a}\,_c\left(\frac{-4 \Delta c'(\Delta) + c(\Delta) (2 + \upeta)}{ c^2(\Delta)  + \Delta}  \right)\,^c\,_d  \gamma^{db}  \Sigma_{,b}  -  C^{a}\,_b  \left(  \gamma^{bc}   \dot{A}_{,c}   +   \dot{\gamma}^{bc}  \frac{1}{2} A_{,c}     \right)
\eeq
With this choice we still retain the freedom to pick the explicit form of $c$, $Y$ and $\mathfrak{c}$.

 As we have shown in appendix~\ref{kappa_details} we have a tree-level solution to the flow equation equation given by \eq{Classical_action} which has an equation of motion proportional to $\E$. We can then evaluate the one-loop flow of the effective action up to terms which vanish for $\E = 0$. Since $S_{,a}$ vanishes when $\E = 0$ the one-loop flow is given by
\beq \label{one_loop0_explicit}
\L \partial_\L S_1[\phi]  = \frac{1}{2} \Tr \left[ G^{-1} \dL G - H^{-1} \dL H   + 2 \frac{\delta \Psi}{\delta \phi}  \right]  
+ \mathcal{O}(\E)  \,,
\eeq 
where it remains to compute the trace on the RHS.
Up to terms which vanish for $\E = 0$ we have that
\beq
G_{ab} = \Pi_{ac} \left( ( 1 + \Delta + \Delta X(\Delta)  \Delta)^2 \right)^c\,_b  +  \Xi_{ac} \left(  (1 + \Delta Y(\Delta)^2     )  \right)^c\,_b
\eeq
and 
\beq
H_{\a\b} = \eta_{\a\c}\left(   \left( 1 + Y( \Delta^{\parallel} )  \Delta^{\parallel}  \right)^2 ( \mathfrak{c}( \Delta^{\parallel} )  +  \Delta^{\parallel}  )                \right)^\c\,_\b 
\eeq
taking a derivative with respect to $\L$ we have
\bea
\dL  G_{ab} = && \Pi_{ac} \left( \frac{2 (c(\Delta )+\Delta ) \left(2 \Delta ^2 c'(\Delta )-\Delta  c(\Delta
   )+c(\Delta )^2\right)}{c(\Delta )^3} \right)^c\,_b \nn
   && +  \Xi_{ac} \left(-2 (\Delta  Y(\Delta )+1) \left(\Delta  \left(2 \Delta  Y'(\Delta )+Y(\Delta
   )\right)-1\right)   )  \right)^c\,_b
\eea
and
\beq
\dL H_{\a\b} = \eta_{\a\c}\left(   -2 \left(\Delta^{\parallel}  \left((\Delta^{\parallel}  Y(\Delta^{\parallel} )+1) \mathfrak{c}'(\Delta^{\parallel})+\Delta^{\parallel 2} Y'(\Delta^{\parallel}
   )-1\right)+\mathfrak{c}(\Delta^{\parallel} ) \left(\Delta^{\parallel 2} Y'\Delta^{\parallel})-\Delta^{\parallel}  Y(\Delta^{\parallel}
   )-2\right)\right)  )                \right)^\c\,_\b 
\eeq

The three operators  which appear under the trace in \eq{one_loop0_explicit} are given by
\beq \label{deltaPsi/deltaphi}
\frac{\delta \Psi}{\delta \phi} =   \left(   \frac{\Delta( -4\Delta c'(\Delta) + c(\Delta) ( 2 + \upeta))    }{2 c(\Delta) ( c^2(\Delta) + \Delta) }         \right)   \Pi  \,,
\eeq

\beq
 G^{-1} \dL G =     \frac{2 \left(2 \Delta ^2 c'(\Delta )-\Delta  c(\Delta )+c(\Delta
   )^2\right)}{c(\Delta ) (c(\Delta )+\Delta )} \Pi    + \frac{2-2 \Delta  \left(2 \Delta  Y'(\Delta )+Y(\Delta )\right)}{\Delta 
   Y(\Delta )+1} \Xi\,,
\eeq
and
\beq
 H^{-1} \dL H  = \frac{2 \left(\mathfrak{c}(\Delta )-\Delta  \mathfrak{c}'(\Delta )\right)}{\mathfrak{c}(\Delta )+\Delta }     +    \frac{2-2 \Delta  \left(2 \Delta  Y'(\Delta )+Y(\Delta )\right)}{\Delta 
   Y(\Delta )+1} 
\eeq
dropping all terms of order $\E$.
Using the identities $\Pi = \delta - \Xi$ and  $\Tr[ f(\Delta) \Xi] = \Tr[ f(\Delta^{\parallel})  ]  + \mathcal{O}(\E)$ we obtain the final result
\beq \label{S_1_flow_exp}
\L \partial_\L S_1 = \Tr[ W(\Delta)]   + \Tr[ W_{\parallel}(\Delta_{\parallel})]  + \mathcal{O}(\E)
\eeq
where the functions $W$ and $W_{\parallel}$ are given by
\beq
 W(z) = \frac{2 z ^2 (c(z)-1) c'(z )+c(z ) \left(c(z)^2-z  c(z )+2 z \right)}{(c(z )+z ) \left(c(z)^2+ z \right)}
\eeq
and 
\beq
W_{\parallel}(z)  =   -  \frac{ \mathfrak{c}(z )-z  \mathfrak{c}'(z)}{\mathfrak{c}(z)+z}  -  W(z)
\eeq
Note that the dependence on $Y$ has cancelled between the terms involving $G$ and $H$.
The scheme independent features of $W(z)$ and $W_{\parallel}(z)$ are that they both go to zero as their argument diverges and that for vanishing momentum
\beq \label{W_uni}
W(0) = 1\,    \,,  \,\,\,\,\,\,\,\,  \,\,\,\,\,\,\,\,      W_{\parallel}(0) = -2   \,.   
\eeq
To find the scheme independent information in the traces we can expand each trace in the local heat kernel expansion and look at the terms that give rise to the logarithmic 
renormalisation of dimensionless couplings. In even dimensions these are given by the terms in the expansion which are proportional to the heat kernel coefficient $B_{\frac{D}{2}}(\Delta)$. Explicitly we have
\beq
\Tr[ W(\Delta)] =  ... + W(0)    \frac{1}{(4 \pi)^{\frac{D}{2}}} B_{\frac{D}{2}}(\Delta) + ... \,, \,\,\,\,\,\,\,\,\,\,\,\,\,\,\,\,   \,\,\,\, \Tr_{\G}[ W_{\parallel} (\Delta_{\parallel})] = ... +  W_{\parallel}(0)   \frac{1}{(4 \pi)^{\frac{D}{2}}} B_{\frac{D}{2}}(\Delta_{\parallel}) + ...  \,,
\eeq
which only depend on the universal numbers  \eq{W_uni}. 
As such we find
\beq \label{One_loop}
\Lambda \partial_{\Lambda} S_1  =  ... +    \frac{1}{(4 \pi)^{\frac{D}{2}}} B_{\frac{D}{2}}(\Delta)      - 2   \frac{1}{(4 \pi)^{\frac{D}{2}}} B_{\frac{D}{2}}(\Delta^{\parallel}) +...   \,,
\eeq
up to scheme dependent terms (i.e. off-shell terms and terms that scale with a power of $\Lambda$) and non-local terms which are not captured by the local heat kernel expansion.

For Yang-Mills theory  
in $D=4$ dimensions the relevant heat kernel coefficients the operators \eq{DeltaYM} and \eq{DeltaghYM} are 
\beq
B_{2}(\Delta) 
=    \frac{5}{3} N \int d^4x F^{\mu\nu}_{i} F_{i\mu\nu}   \,,
\eeq
and
\beq
B_{2}(\Delta^{\parallel}) = - \frac{1}{12} N   \int d^4x F^{\mu\nu}_{i} F_{i\mu\nu}   \,.
\eeq
Inserting the heat kernel coefficients into \eq{One_loop} we find that the Yang-Mills action is renormalised by  
 \beq
 \Lambda \partial_{\Lambda} \frac{1}{4\cc^2} \int d^4x F_{i\mu\nu} F^{i\mu\nu}  =   \frac{1}{(4 \pi)^{2}}   \frac{11N}{6}  \int d^4x   F_{i\mu\nu} F^{i\mu\nu} \,,
\eeq
and hence we find the expected one-loop beta function
\beq
\beta_{\cc^2} = - \frac{ 1}{(4\pi)^2} \frac{22N}{3}  \, \cc^4 \,.
\eeq

For pure gravity up to terms linear in the Einstein Equations the operators $\Delta$ are given by
\beq
\Delta \delta g_{\mu\nu} =  \Lambda^{-2} (-\nabla^2 - 2 {\rm Rie} )\delta g_{\mu\nu}  \,,\,\,\,\,\,\,\,\,    \Delta^{\parallel} \delta v^{\mu} =  \Lambda^{-2} \left(-\nabla^2 - \frac{2}{D-2} \bar{\lambda} \right)\delta v^{\mu}  \,,
\eeq
where ${\rm Rie}\, \delta g_{\mu\nu} = R^{\rho}\,_{\mu}\,^{\sigma}\,_{\nu}  \delta g_{\sigma \rho}$. In $D=4$ dimensions these operators have the heat kernel coefficients
\beq
B_{2}(\Delta) =  \int d^4x \sqrt{\g}   \left( \frac{19}{18} R_{\mu\nu\rho\sigma} R^{\mu\nu\rho\sigma} - \frac{2}{3} \bar{\lambda}^2 \right)\,,
\eeq
\beq
B_{2}((\Delta^{\parallel}) =  \int d^4x \sqrt{\g}  \left(  - \frac{11}{180}  R_{\mu\nu\rho\sigma} R^{\mu\nu\rho\sigma} + \frac{82}{15}  \bar{\lambda}^2 \right)\,,
\eeq
and thus from \eq{One_loop} we obtain the well known result \cite{Christensen:1979iy}
\beq  \label{One_loop_GR}
\Lambda \partial_{\Lambda} S_1 =   ...  +  \frac{1}{(4 \pi)^{2}}  \int d^4x \sqrt{\g}   \left( \frac{53}{45} R_{\mu\nu\rho\sigma} R^{\mu\nu\rho\sigma}-\frac{58}{5} \bar{\lambda}^2 \right) + ...\,,
\eeq
obtained here for the first time without fixing the gauge. We note that the more general result applies for more general theories \eq{One_loop} such as gravity coupled to matter.

\section{Non-perturbative running of the gauge coupling}
\label{Non-Pert_g}

In the last section we showed how one-loop results can be recovered from our flow equation. However the real advantage of having an exact flow equation is that one is not limited to perturbation theory and can instead use non-perturbative approximations. To exemplify this let us consider a truncation of the effective action to be for the form 
\beq \label{Truncation}
S = I + \frac{1}{2} A_{,a} \gamma^{ab} A_{,b}
\eeq
which involves the cutoff term needed to regularise the theory along with the action $I$.
 We then aim to compute the flow of the Yang-Mills gauge coupling in this truncation. For this we can ignore terms which are proportional to the equation of motion $ \E_{a}$ and thus all we need is the trace \eq{one_loop0_explicit} we evaluated in the previous section but now retain the terms which involve the anomalous dimension $\upeta$.

\subsection{Incorporating the  anomalous dimension}
\label{anom_dim_g}
Here the anomalous dimension is given by
\beq
\upeta   =  {\rm g}^{-2} \L \partial_\L  {\rm g}^2
\eeq
since $G \propto 1/ {\rm g}^2$ and $H \propto  1/{\rm g}^2$ we have that
\beq
 G^{-1} \dL G =     \frac{2 \left(2 \Delta ^2 c'(\Delta )-\Delta  c(\Delta )+c(\Delta
   )^2\right)}{c(\Delta ) (c(\Delta )+\Delta )} \Pi    + \frac{2-2 \Delta  \left(2 \Delta  Y'(\Delta )+Y(\Delta )\right)}{\Delta 
   Y(\Delta )+1} \Xi\  - \upeta (\Pi + \Xi) ,
\eeq
and
\beq
\left( H^{-1} \dL H  \right)^{\a}\,_\b =  \left( \frac{2 \left(\mathfrak{c}(\Delta )-\Delta  \mathfrak{c}'(\Delta )\right)}{\mathfrak{c}(\Delta )+\Delta }     +    \frac{2-2 \Delta  \left(2 \Delta  Y'(\Delta )+Y(\Delta )\right)}{\Delta 
   Y(\Delta )+1}   - \upeta \right)\,^\a\,_\b
\eeq
whereas the functional derivative of $\Psi$ is given by \eq{deltaPsi/deltaphi} with the term proportional to $\upeta$ retained. 
The flow then takes the form
\beq
\L \partial_\L S = \Tr[ W(\Delta)  + \upeta \tilde{W}(\Delta)  ]   + \Tr[ W_{\parallel}(\Delta_{\parallel})  + \upeta \tilde{W}_{\parallel}(\Delta_{\parallel}) ]  + \mathcal{O}(\E)
\eeq
where the two new functions are given by
\beq
\tilde{W}(z)= - \tilde{W}_{\parallel}(z)  =  -\frac{c(z)^2}{2 \left(c(z)^2+z\right)} 
\eeq
As we see the traces proportional to $\upeta$ are regulated. 

To extract the running of the gauge coupling we need only the term proportional to $B_2 \propto \int d^4 \tr F_{\mu\nu} F^{\mu\nu}$ which gives
\bea \label{Gauge_running_NP}
\L \partial_\L   \frac{1}{4 {\rm g}^2} \int d^4 \tr F_{\mu\nu} F^{\mu\nu} &=&  \frac{1}{(4 \pi)^{2}}   \left( 1 - \frac{\upeta}{2} \right) B_2(\Delta)   -    \frac{1}{(4 \pi)^{2}}    \left( 2 - \frac{\upeta}{2} \right)   B_2(\Delta_{\parallel})  \\   
&=&   \frac{N}{(4 \pi)^{2}}   \left( \frac{11}{6} - \frac{7}{8} \upeta \right)    \int d^4x \tr F^{\mu\nu} F_{\mu\nu}   
\eea
This leads to the non-perturbative beta function
\beq
\beta_{\cc^2} = -\frac{44 \cc^4 N }{3 \left(32 \pi ^2-7 \cc^2 N \right)} \,.
\eeq
Let us now make several comments on this result. Firstly we have clearly made a drastic approximation by the truncation \eq{Truncation} which amounts to setting $\S =0$. However since our equation is background independent we can systematically improve upon this result by adding further terms to $\S$ with increasing numbers of covariant derivatives without having to additionally take into account additional dependence on a background field or ghost fields. The fact that we have made an approximation is apparent from the fact that at two-loop order we have
\beq \label{two_loop_beta_gauge}
\beta_{\cc^2} =  - \frac{ 1}{(4\pi)^2} \frac{22}{3}  N \, \cc^4  -   \frac{ 1}{(4\pi)^4} N^2  \frac{77}{3}  \cc^6 + ... \,.
\eeq
which differs from the universal two-loop result. The two loop-coefficient in \eq{two_loop_beta_gauge} agrees with that found in \cite{Gies:2002af} up to scheme dependent corrections which are absent in here due to our truncation.
 To obtain the correct result we would need to expand our truncation \eq{Truncation}.

The result has a very straightforward interpretation. Recall that the breaking of scale invariance of a classically scale invariant theory has its origin in the fact that the path integral measure is not scale invariant. We see this breaking explicitly in the continuum limit $\L \to \infty$ since $ G_{ab} \to \gamma_{ab} \propto \Lambda^2$ and $  H_{\a\b}  \to \eta_{\alpha \beta} \propto \L^4$. However we should also take into account that the measure depends also on the gauge coupling which then depends on $\L$ through renormalisation. The breaking of scale invariance due to the continuum measure is therefore characterised completely by 
the `anomalous scaling dimensions' $\upeta_\gamma$  and $\upeta_\eta$ given by
\beq
\L \partial_\L \gamma_{ab} = - \upeta_{\gamma}  \gamma_{ab}  \,,    \,\,\,\,\,\,\,\,\,\,\,  \L \partial_\L \eta_{ab} = - \upeta_{\eta}  \eta_{ab} 
\eeq
which given explicitly by 
\beq
\upeta_{\gamma}  = -2 + \upeta\,, \,\,\,\,\,\,\,\,\,\,\,    \upeta_{\eta}  = -4 + \upeta
\eeq
We can then rewrite \eq{Gauge_running_NP} as
\bea \label{Gauge_running_NP}
\L \partial_\L   \frac{1}{4 {\rm g}^2} \int d^4 \tr F_{\mu\nu} F^{\mu\nu} &=&   - \frac{1}{2} \upeta_{\gamma}  B_2(\Delta)+ \frac{1}{2} \upeta_{\eta}     B_2(\Delta_{\parallel})     
\eea
and thus  within our approximation we see that the beta function is characterised by the breaking of scale invariance due to the measure.

\subsection{Comparison to the background field approximation}
\label{compare_to_BFA}
The running \eq{Gauge_running_NP} resembles the running found obtained from the flow of the effective average action in background field approximation \cite{Reuter1994181} and agrees with the beta function found in \cite{Wetterich:2017aoy}. In the effective average action approach one has a background field as well as the dynamical one and  the gauge field is split into the background $\bar{A}_{\mu}$ and a fluctuation $a_\mu$ such that
\beq
A_\mu = \bar{A}_{\mu} +  Z_a^{1/2} \cc a_\mu
\eeq
where $Z_a^{1/2}$ is the wavefunction renormalisation of the fluctuation. In addition to the gauge field one also has  Faddeev-Popovv ghosts  which also have a wavefunction $Z_{\rm gh}$ if the ghosts are rescaled as
\beq
C_{\rm gh} \to  \cc^{1/2} Z^{1/2}_{\rm gh} C_{\rm gh}\,,  \,\,\,\,\,\,\,\,\,\,\,   \bar{C}_{\rm gh} \to  \cc^{1/2} Z^{1/2}_{\rm gh} \bar{C}_{\rm gh}
\eeq
The background field approximation consists of assuming the separate dependence on $\bar{A}_\mu$ comes only from the gauge fixing and extracting the running of $\cc$ from of the effective action with $a_\mu= 0$. 
The result is given by
\beq
\L \partial_\L   \frac{1}{4 {\rm g}^2} \int d^4 \tr F_{\mu\nu} F^{\mu\nu} =    \left( 1 - \frac{\upeta_a}{2} \right)   B_2(\Delta)     -     \left( 2 -\upeta_{\rm gh} \right)   B_2(\Delta_{\parallel}   ) 
\eeq
In order to close this approximation one has to fix $\upeta_a= - Z^{-1} \L \partial_\L Z_{a}$ and $\upeta_{\rm gh} = - Z^{-1}_{\rm gh} \L \partial_\L Z_{\rm gh}$.
Evidently we recover our result if
\beq \label{Gauge_Invariant_etas}
\upeta_{a} = \upeta\,, \,\,\,\,\,\,\,   \upeta_{\rm gh} = \frac{1}{2} \upeta
\eeq
which corresponds to not rescaling the fields by taking  $Z_a^{1/2} \cc=1$   and $ \cc^{1/2} Z^{1/2}_{\rm gh} =1$.

One then understands that disagreement between the beta function \eq{Gauge_running_NP} and the one found in \cite{Reuter1994181} is due to the choice to neglect the ghost anomalous dimension in the latter by setting $\eta_{\rm gh} =0$.
However, to close the approximation consistently in the effective average action approach one can find the flow of the of the two point functions for $a_\mu$ and the ghosts to extract $\upeta_{a}$ and $\upeta_{\rm gh}$ \cite{Codello:2013wxa}. 
Within a background dependent scheme this `extra' calculation is needed for consistency to take into account that the flow equation depends on the ghosts and the background field in addition to the gauge fields. 
In our background independent scheme there is no separate  dependencies on a background and we have no ghosts and thus we close the approximation consistently without any extra calculation.
To say this in another way, if we demand that our approximation in either scheme should include all terms with up two derivatives, in the background independent scheme we have closed this approximation by including only the single gauge coupling whereas in a background dependent approach there are more independent couplings which must be included to close the approximation to this order.

\section{Conclusions}
\label{conc}
In this work we have presented a manifestly background independent formalism which can be applied to both quantum gravity and gauge theories to address perturbative and non-perturbative questions within these fields. This approach is based on a consistent gauge invariant regularisation scheme which cuts off momentum integrals at the scale $\L$. For Yang-Mills theories gauge invariant regularisation schemes have a long history culminating in consistent schemes which avoid over lapping divergencies \cite{Asorey:1989ha,Bakeyev:1996is}. The essential features of these schemes are shared by our scheme which also provides a diffeomorphism invariant regularisation for quantum gravity. In our scheme the necessary Pauli-Villars regularisation consists simply of functional determinants in the path integral measure without
needing to include extra fields to regularise the theory. In this way we avoid including extra fields in the action which must be included in the regularisation schemes \cite{Morris:2000fs,Morris:2000jj}  used in previous manifestly gauge invariant ERGs. 
The reason we can avoid including additional interacting fields due to the Pauli-Villars operators diverging sufficiently slowly relative to the rate at which the physical propagators decrease for high momentum. Without this property we would need to include  PV fields with self-interactions to cancel additional over lapping divergences.

 Based on our regularisation we have derived an ERG flow equation which tells us how the theory is renormalised as we integrate out high momentum degrees of freedom. The flow equation governs the $\L$ dependence of the Wilsonian effective action $S[\phi]$. Solving the flow for $\infty \leq \L \leq 0$ we have a microscopic theory at $\L \to \infty$ while in the limit $\L \to 0$    we compute the functional integral using \eq{ZL0}. Expectation values of  gauge invariant operators  can be obtained from by including gauge invariant source terms in the action \cite{Rosten:2006pd,Rosten:2006qx}.   The flow equation therefore provides a formalism to solve a quantum field theories without the need of a fixed background geometry. Furthermore, since we do not have to fix the gauge, the flow equation does not suffer from the Gribov ambiguity \cite{1978NuPhB.139....1G}.

In comparison to the effective average action, which depends separately on a non-dynamical background field, our construction avoids any additional background field dependence. Thus the flow equation is solved for a gauge invariant action $S[\phi]$ which involves only the physical fields which appear in the classical action. As we pointed out in the introduction the advantage of this approach is that we do not have a proliferation of couplings encountered if the action would depended separately  on a background field.  Thus within this formalism the Ward identities \eq{Ward}, which can be exploited when expending the flow equation around a background, take there `classical form' even at the quantum level. Alternatively if we expand the flow equation in a derivative expansion the number of independent couplings at each order is vastly reduced compared to background dependent flow equations.  Consequently, although the flow equation is apparently more complex than background dependent flow equations, this complexity is massively reduced by the background independence and manifest gauge invariance of the formalism.

\section*{Acknowledgements}
This work has benefited from discussions with  Tim Morris, Jan Pawlowski, Robertio Percacci and Oliver Rosten.
In the earlier stages of this work the author was supported by the  European Research Council grant ERC-AdG-290623.

\begin{appendix}

\section{On-shell identities}
\label{OnShell_id}

One can make us of several identities when the field equation \eq{EofM} hold which follow from noting that then since $I_{,b} K^b_{\a} = 0$ we have 
\beq
I_{,ba} K^b_{\a}  = - \E_{b} K^b_{\a,a}
\eeq
and thus
\beq
I_{;ab} K^b_{\a} =   - \E_{b} K^b_{\a,a}   - \E_{c} \Gamma^c_{ab} K^{b}_{\a}    
\eeq
 vanishes for a solution to $\E_{a} = 0$. One then has that
 \beq
 \Delta^a\,_b K^b_{\a} = K^a_{\b}  (\Delta^\parallel)^\b\,_{\a}  + \mathcal{O}(\E)
 \eeq
 and
 \beq
 \Delta^a\,_c \Pi^{c}\,_b =   (\Delta^\perp)^{a}\,_b  + \mathcal{O}(\E)
 \eeq
Repeated uses of these identities also imply that 
\beq
   \left( \Delta W(\Delta) \right)^a\,_{c}  \Xi^c\,_b  =  K^a_{\b}  W(\Delta^\parallel)^\b\,_{\a}  K^{\a}_{b}  + \mathcal{O}(\E)
\eeq
 and
 \beq
  \left(  W(\Delta) \right)^a\,_{c}  \Pi^c\,_b  = W(\Delta^{\perp}) ^a\,_{b}  + \mathcal{O}(\E)
 \eeq
 for suitably behaved functions $W(z)$.

\section{$\E$-expansion}
\label{kappa_details}
Here we will to show that  \eq{Classical_action} is a solution to the classical flow equation \eq{tree_level_flow} with $\kappa$ given by \eq{kappa} up to order $\E^2$.
From \eq{Classical_action} and \eq{Sigma} we have
\beq
\Sigma = - \frac{1}{2} \E_{a} X^{ab} \E_{b}  + \mathcal{O}(\E^3)   +  \mathcal{O}(\hbar) 
\eeq
additionally we have the equations
\bea
\Sigma_{,a} = -   \E_{c} X^{c}\,_b  \Delta^b\,_a  + \mathcal{O}(\E^2) \\
\hat{S}_{,a} = \E_{c} \left( (1 + X \Delta)(1+ \Delta + \Delta^2 X) \right)^{c}\,_{a}   + \mathcal{O}(\E^2) \\
 \dot{\gamma}^{ab} = (-2 + \upeta) \gamma^{ab}\\
\dL \E_{a}  = - \upeta  \E_{a}\\
\dot{\Delta}^a\,_b = - 2  \Delta^{a}\,_b\\
A_{,a} = \E_{b} (1 + \Delta X )^b\,_a   + \mathcal{O}(\E^2) \\
\dot{A}_{,a} =  \E_{,b} ( -2 \Delta X  +  \Delta \dot{X} - \upeta(1 + \Delta X)    )^b\,_a  + \mathcal{O}(\E^2)
\eea
Inserting these into \eq{tree_level_flow} we obtain
\bea \label{tree_level_flow2}
&& - \upeta I  + \frac{1}{2} \E_{c} \left(1 + \Delta( 1 + X \Delta)^2 \right)^{c}\,_{a}  \K^{ab}  (\Delta X)^{d}\,_b   \E_{d}   \nn &&+  \E_{c} \left(  \frac{\Delta X}{1 + \Delta + \Delta X \Delta}    \right)^{ca} \left( \dot{X} \Delta - 1 - 3 X \Delta - \frac{\upeta}{2}( 1 + X \Delta)       \right)^d\,_a \E_{d}    = 0   + \mathcal{O}(\E^2) + \mathcal{O}(\hbar) \,.
\eea
Now we can take $\upeta \propto \hbar$ such that $\upeta$ is a quantum correction and we can drop the first term. The two terms other terms are of the form $\E( ... ) \E$ and thus we can solve for $\K$ and hence for $\kappa$.
It is convenient to express $\kappa$ in terms of $c(z)$ to arrive at the explicit expression \eq{kappa}.
One can compute higher orders in $\E$ which determine $\S$ systematically by requiring that $\S$ vanishes for $\L \to \infty$.

\section{Gauge fixing}
\label{GFsec}
\subsection{Gauge independence}
\label{GF}
Let us show that the gauge fixed path integral is independent of the gauge condition $\chi$.
To achieve this it is sufficient to show that a change in the gauge is equal to a field redefinition
\beq \label{Gauge_indep}
(\delta_\chi  + \mathcal{L}_{V})  \sqrt{\det G[\phi] }  \frac{1}{\sqrt{\det H[\phi]} }  e^{-S[\phi]  -\frac{1}{2} \chi^\a Y_{\a\b} \chi^\b } (\det Q)     =   0     
\eeq
where $\delta_\chi$ denotes a variation of the gauge condition.
In particular \eq{Gauge_indep} holds for 
\beq
V^a[\phi] =  - K^a_{\a}[\phi] (Q^{-1}[\phi])^\a\,_\b \delta \chi^\b[\phi] 
\eeq
as can be shown by direct computation. Intermediate  steps yeild
\beq
(\delta_\chi  + \mathcal{L}_{V})  e^{-S[\phi] -\frac{1}{2} \chi^\a Y_{\a\b} \chi^\b} = 0 \,,
\eeq
\bea
(\delta_\chi  + \mathcal{L}_{V})  \sqrt{\det G[\phi] } & =&   -   \sqrt{\det G } K^a_{\a}   \left ( (Q^{-1})^\a\,_{\b,a} \delta \chi^\b + (Q^{-1})^\a\,_{\b} \delta \chi^\b_{,a} \right) \,,
\eea
\bea
(\delta_\chi  + \mathcal{L}_{V})  \det Q
&=&    \det Q\left(   (Q^{-1})^\a\,_{\b} \delta \chi^\b_{,a} K^a_{\a}   + (Q^{-1})^\a\,_{\b,a}   \delta\chi^\b K^a_{\a}   +   f^{\a}\,_{\a\c}  (Q^{-1})^\c\,_{\delta}\delta\chi^\delta \right) \,,
\eea

\beq
(\delta_\chi  + \mathcal{L}_{V})  \frac{1}{\sqrt{\det H_{\a\b}[\phi]}}  = -  \frac{1}{\sqrt{\det H_{\a\b}[\phi]}}   f^{\a}\,_{\a\c} (Q^{-1})^\c\,_{\delta}\delta\chi^\delta \,,
\eeq
from which we see that all terms cancel.

\subsection{Gauge independent flow}
\label{GFflow}
Using \eq{Gauge_indep}  allows us show that the flow equation \eq{The_Flow_Equation} can be derived also if we fix the gauge.
This follows by noting that the additional terms which are generated due to the gauge fixing terms are equivalent to a change in the gauge fixing condition
\beq \label{chi_shift}
\chi^\a \to \chi^a +  \chi^\a_{,a} \Psi^a  \L^{-1} \delta \L
\eeq
and since the path integral is independent of the gauge fixing condition the flow equation is still valid.  Explicitly we just have to shift $\Psi^a \to \Psi^a   - K^a_{\a}[\phi] (Q^{-1}[\phi])^\a\,_\b \chi^\b_{,b} \Psi^b$ such that all the gauge dependent terms cancel.
To see this explicitly note that under  \eq{chi_shift} we have
\bea
S_{\rm BRST} \to&& S_{\rm BRST}  + i B^{\a} \bar{Y}_{\a\b}   \chi^\b_{,a} \Psi^a  \L^{-1} \delta \L   + \bar{C}^\a Y_{\a\b}   (\chi^\b_{,a} \Psi^a)_{,b} K^a_{\c} C^{\c} \L^{-1} \delta \L \nn
&=&  S_{\rm BRST}  +   \left( i B^{\a} \bar{Y}_{\a\b}   \chi^\b_{,a} \Psi^a  \L^{-1}   + \bar{C}^\a Y_{\a\b}   \chi^\b_{,ab} \Psi^a K^b_{\c} C^{\c}  + \bar{C}^\a Y_{\a\b}   \chi^\b_{,a} \Psi^b K^a_{\c,b} C^{\c} \right) \L^{-1} \delta \L \nn
&=&  S_{\rm BRST}  +    S_{\rm BRST, a} \Psi^a \L^{-1} \delta \L\,.
\eea
Thus
\beq
\mathcal{L}_{\Psi} e^{ - S_{\rm BRST}} = \delta_\chi  e^{ - S_{\rm BRST}}\,  \,\,\,  {\rm for }  \,\,\,   \delta \chi^\a =  \chi^\a_{,a} \Psi^a  \L^{-1} \delta \L
\eeq

\end{appendix}

\bibliography{myrefs,ASreferences}

\end{document}